\documentclass[lettersize,journal]{IEEEtran}
\usepackage{amsmath,amsfonts}
\usepackage[ruled]{algorithm2e}%\usepackage{algorithm}
\usepackage{array}
\usepackage[caption=false,font=normalsize,labelfont=sf,textfont=sf]{subfig}
\usepackage{textcomp}
\usepackage{stfloats}
\usepackage{url,lipsum}
\usepackage{verbatim}
\usepackage{graphicx}
\usepackage{cite}
\usepackage{mathrsfs}
\usepackage{amsmath}
\usepackage{booktabs}
\usepackage{multirow,url}
\usepackage{bm}
\usepackage[table]{xcolor}
\usepackage{bbding}
\usepackage{diagbox}
\usepackage{color}  % To include comments in color

\definecolor{oblue}{rgb}{0.086,0.439,0.710}

\begin{document}

\title{Integrating Plug-and-Play Data Priors with Weighted Prediction Error for Speech Dereverberation}

\author{Ziye Yang,~\IEEEmembership{Student Member,~IEEE}, \, Wenxing Yang,\, Kai Xie, \,Jie Chen,~\IEEEmembership{Senior Member,~IEEE}
        % <-this % stops a space
\thanks{A preliminary version  has been accepted for European Signal Processing Conference (EUSICPO 2023).  This version provides detailed algorithm description and comprehensive experiments to illustrate the method. }
\thanks{The authors are with Center of Intelligent Acoustics and Immersive Communications, Northwestern Polytechnical University, China.}% <-this % stops a space
\vspace{-5mm}}

% The paper headers
\markboth{Journal of \LaTeX\ Class Files,~Vol.~14, No.~8, August~2021}%
{Shell \MakeLowercase{\textit{et al.}}: A Sample Article Using IEEEtran.cls for IEEE Journals}

%\IEEEpubid{0000--0000/00\$00.00~\copyright~2021 IEEE}
% Remember, if you use this you must call \IEEEpubidadjcol in the second
% column for its text to clear the IEEEpubid mark.
% ,\, Kai Xie,~\IEEEmembership{Student Member,~IEEE},

\maketitle

\begin{abstract}
Speech dereverberation aims to alleviate the detrimental effects of late-reverberant components. While the weighted prediction error (WPE) method has shown superior performance in dereverberation, there is still room for further improvement in terms of performance and robustness in complex and noisy environments. Recent research has highlighted the effectiveness of integrating physics-based and data-driven methods, enhancing the performance of various signal processing tasks while maintaining interpretability. Motivated by these advancements, this paper presents a novel dereverberation framework, which incorporates data-driven methods for capturing speech priors within the WPE framework. The plug-and-play strategy (PnP), specifically the regularization by denoising (RED) strategy, is utilized to incorporate speech prior information learnt from data during the optimization problem solving iterations. Experimental results validate the effectiveness of the proposed approach.

%Speech dereverberation aims to mitigate the impact of late-reverberant components. As a typical  approach to dereverberation, the weighted prediction error (WPE) method has shown its superior performance, however it is still possible to further improve its performance and robustness by incorporating sophisticated speech priors.  Recent research demonstrates  that the integration of physics-based and data-driven methods can improve the performance of various signal processing tasks while maintaining the interpretability of the problem solving process.  Motivated by the relevant progress, this paper presents a novel dereverberation framework, termed PnPWPE, that incorporates the data-driven method for speech prior capturing for WPE. The plug-and-play strategy (PnP), specifically the regularization by denoising (RED) strategy, is used to incorporate speech prior information during the alternating direction method of multipliers (ADMM) solving iterations by plugging in a pre-trained speech denoiser. Experimental results demonstrate the effectiveness of the proposed method.
%{\footnote{Demo results are available at \url{ https://github.com/yzy-nwpu/PNP_WPE-for-speech_dereverb}.}}
\end{abstract}

\begin{IEEEkeywords}
Speech dereverberation, the weighted prediction error method, data-driven method, learnt speech priors.
\end{IEEEkeywords}

\section{Introduction}

Speech signals captured in enclosed rooms by far-field microphones are inevitably affected by energy consumption and reflections from the room's walls, ceilings, and other rigid objects during propagation. These effects result in delayed and attenuated copies of the source signal, known as speech reverberation. Reverberation causes degradation in the quality of the speech of interest, impacting various higher-level speech applications, including automatic speech recognition systems, speaker identification systems and teleconference systems \cite{chetupalli2019late}. Despite the challenge of differentiating the direct speech from its reverberation components, speech dereverberation has garnered significant attention.
%\cite{williamson2017time, schwartz2016expectation, inoue2019joint}.

% \begin{figure}[!t]
% \centering
%      \includegraphics[width=0.5\textwidth]{img/rirs.pdf}
%  \caption{(a)~is an example of scenario containing a speaker, a microphone array and ambient noise. (b)~is an example of room impulse response~(RIR). Please refer to~\eqref{eq:signalpnp} for the explanation of variables in the figure.}
%  \label{fig:rir}
%\end{figure}

Reverberation components can be categorized into early and late components based on the arrival time at the microphone array, which is influenced by factors such as room size, distance from the source to the microphones, and damping properties of the environment's surfaces \cite{kothapally2022skipconvgan}. %\cmag{In Fig.\ref{fig:rir}, an example scenario with a speaker, a microphone array, and ambient noise is depicted. The initial short period of near-zero amplitude represents the direct-path propagation from the source to the microphones (indicated by the red line in Fig.\ref{fig:rir}(a)), and the first peak corresponds to the direct-path component.}
Typically, the subsequent 50 ms of the impulse response is considered as the early-reverberant components \cite{kuttruff2016room}, while the remaining duration is referred to as the late-reverberant components. Research has shown that early-reverberant components are not perceptible to human hearing and improve the quality of the direct-path component~\cite{naylor2010speech}. Therefore, dereverberation techniques aim to eliminate the late-reverberant components while preserving the early-reverberant ones and the direct-path signal, as only the late-reverberant components have a detrimental effect on speech intelligibility and quality \cite{schmid2014variational, kinoshita2013reverb}.

Considerable efforts have been devoted to devising speech dereverberation methods, which can be primarily categorized into conventional signal processing-based methods (i.e., physics-based methods) and data-driven algorithms. The former addresses the dereverberation problem based on the speech convolution model, with typical methods including acoustic channel equalization-based approaches \cite{kodrasi2016joint}, suppression-based methods \cite{kodrasi2018analysis},
%braun2018evaluation}
beamforming-based methods \cite{huang2020simple}, and linear prediction-based methods \cite{nakatani2008speech}. These methods often model the acoustic transfer function as an auto-regressive or convolutive problem \cite{nakatani2010speech, jukic2014speech, schwartz2014online}, with the spectral coefficients of clean speech being modeled as a Gaussian or Laplacian distribution. Dereverberation is then performed by maximum likelihood estimation of unknown parameters \cite{kodrasi2016joint}. When multiple microphones are available, spatial information can be leveraged to filter out signals arriving from undesired directions using beamforming algorithms \cite{wang2020deep}. For example, the work in \cite{braun2018evaluation} decomposes the multi-channel Wiener Filter into a minimum variance distortionless response and a single-channel post filter, enabling noise reduction and dereverberation in a two-stage approach. Among the numerous model-based dereverberation techniques, multichannel linear prediction (MCLP) methods, which estimate and subtract the late-reverberant components from the observed signal, show significant promise. One such method is the weighted prediction error (WPE) method \cite{nakatani2010speech}, which has demonstrated its effectiveness in dereverberation. To further enhance performance, works such as \cite{jukic2015multi} and \cite{witkowski2021split} propose leveraging speech sparsity in the time-frequency domain to incorporate an additional prior on the unknown variance. Despite the clear physical interpretation of these model-based methods, their performance and robustness are limited as they do not fully exploit the inherent priors of speech signal structures.

In contrast, existing data-driven methods predominantly rely on deep learning and treat speech dereverberation as a supervised learning problem \cite{wang2018supervised}. With large amounts of training data and increasing computational resources, data-driven methods have achieved state-of-the-art performance and have become a significant focus in the speech processing community. Initially, deep neural networks (DNNs) were trained to predict real-valued masks or magnitudes of the direct signal in the magnitude domain, with phases being utilized solely in the speech reconstruction stage \cite{han2015learning, mimura2015speech}. To better leverage spectral information, an extension was proposed, wherein the magnitude-domain masking and mapping method was adapted to the complex domain. This approach predicted the real and imaginary components of the direct-path signal from the received reverberant signals \cite{wang2020deep}. Subsequently, novel neural network architectures such as self-attention and recurrent networks were employed for end-to-end modeling in other front-end speech processing tasks \cite{luo2020dual, borgstrom2020speech}. Building on these advances, a monaural algorithm utilizing temporal convolutional networks with self-attention was proposed for end-to-end dereverberation in the time domain \cite{zhao2020monaural}. Data-driven methods, whether applied in the time-frequency or time domain, aim to learn a mapping function from input reverberant signals to output clear speech, incorporating speech priors into the network parameters. The powerful non-linear modeling capability of DNNs enables these methods to extract high-level features, leading to promising performance. However, they often neglect the explicit exploitation of the linear convolutional structure of reverberation \cite{wang2021convolutive}, and the black-box nature of DNNs limits the physical interpretability of the entire speech recovery process.

Due to the respective merits and drawbacks of physics-based methods and data-driven methods, their integration has garnered significant attention in the signal processing community \cite{shlezinger2022model, Bai2020, monga2022unrolling}. This integration benefits from both tractable mathematical modeling and highly-parameterized generic mappings tuned with large data. One promising strategy involves the plug-and-play technique (PnP), which incorporates a deep denoising algorithm as a module into the optimization iterations to capture data priors. The PnP strategy has been successfully investigated in various tasks, including inverse problems in image processing, which shares similar inherent problem formulations with the dereverberation problem. In these tasks, incorporating a denoiser implicitly characterizes image priors, enabling physics-based methods to solve inverse problems more effectively \cite{chan2016plug, zhang2021plug, chen2023integration, wang2021hyperspectral, zhao2021plug, ahmad2020plug}.

%For instance, one direction, which is called DNN-WPE~\cite{kinoshita2017neural,heymann2019joint}, adopts a straightforward but quite effective idea of incorporating a DNN-based spectrum estimator into the vanilla WPE. Whereas it is an interesting attempt to take advantage of DNN to boost the dereverberation performance, these methods simply replace the power spectral density (PSD) module of WPE rather than reconstruct the original problem to achieve the purpose of combining data-driven and physics-based methods, which do not fundamentally promote the framework of WPE by extracting structural information of speech from data.

Inspired by these advances, we propose a framework for speech dereverberation that benefits from both physics-based models and data priors. In this work, we aim to incorporate data priors into the WPE framework from the perspective of formulating the optimization problem. We maintain the interpretability of the problem-solving process while integrating speech prior information learned from data. Specifically, we formulate the prediction error minimization problem of WPE with an additional regularizer that is not explicitly handcrafted. Unlike vanilla WPE and its extensions \cite{jukic2015multi, witkowski2021split, kinoshita2017neural, heymann2019joint}, which do not consider sophisticated speech priors, we propose integrating speech prior information by employing the plug-and-play strategy. Specifically, we employ the regularization by denoising (RED) strategy, which is a promising variant of PnP~\cite{romano2017little}. We incorporate a pre-trained speech denoiser into the optimization iterations, and the proposed framework is referred to as PnPWPE. This approach effectively splits the speech processing problem into two distinct components: a model-based solving part and a data-prior capture part. By incorporating such a denoiser for the latter part, our framework harnesses the power of data-driven learning through denoiser training, eliminating the reliance on task-dependent data. This inherent flexibility makes our approach extendable for other speech processing tasks.

% using the alternating direction method of multipliers (ADMM) \cite{boyd2011distributed}, a widely used variable splitting technique with excellent convergence properties. We refer to the proposed framework as PnPWPE.

%Among the numerous model-based dereverberation techniques, the class of multichannel linear prediction~(MCLP)~methods, which first estimate the late-reverberant components and then subtract them from the observed signal, are, to our knowledge, the most promising. Implementing such a principle, the weighted prediction error (WPE) method~\cite{nakatani2010speech,yoshioka2012making} in particular shows its effectiveness in dereverberation. However, the vanilla WPE method does not take into account sophisticated speech priors, so it is reasonable to expect that the performance of WPE can be notably enhanced by inserting speech prior information learnt from data.

%adopts a straightforward but quite effective idea of incorporating a DNN-based spectrum estimator into the vanilla WPE. Whereas it is an interesting attempt to take advantage of DNN to boost the dereverberation performance, these methods simply replace the power spectral density (PSD) module of WPE rather than reconstruct the original problem to achieve the purpose of combining data-driven and physics-based methods, which do not fundamentally promote the framework of WPE.

%To this end, we reformulate the prediction error minimization problem of WPE with an additional regularizer that is not explicitly handcrafted.

%The rest of this paper is organized as follows. We

\noindent{\textbf{Notation}}.  Normal font letters $x$  and $X$ denote scalars, and boldface small letters $\mathbf{x}$ denote column vectors. Boldface capital letters $\mathbf{X}$ represent matrices, and {the} operator $(\cdot)^{\top}$ and $(\cdot)^{\rm H}$ denote matrix transpose and conjugate transpose respectively. $\text{col}\{\cdots\}$ concatenates its vector arguments.

\section{Signal model and vanilla WPE method}\label{sec:sig_review}
In this section, we first present the signal model in the reverberant scenario, followed by a description of the vanilla WPE method.

\subsection{Signal model for reverberant scenarios}\label{sec:sig}
Consider the scenario where a distant microphone array with $Q$ channels captures the convolved speech. In time domain, the signal model of the $q$-th channel, indicating the relationship between the received reverberant speech $b_q[t]$ and the target source signal ${s}[t]$, can be represented by
\begin{equation}\label{eq:signal}
{b}_{q}[t] =  s[t] \ast h_{q}[t]
\end{equation}
where $t$ indexes discrete time, $\ast$ denotes the linear convolution, and $h_{q}[t]$ is the acoustic impulse responses~(AIRs) between the source and microphone. Equivalently, we can write~\eqref{eq:signal} in vector form
\begin{equation}\label{eq:signalv}
{b}_{q}[t] = \textbf{h}_q ^{\top}\textbf{s}[t],
\end{equation}
with $\textbf{h}_q = \big(h_{q,0},\cdots,h_{q,L-1}\big)^\top$, and $L$ being the order of AIR and $\textbf{s}[t] = \big(s[t],\cdots, s[t-L+1]\big)^\top$ is the speech signal vector.

If the AIRs i.e.,~$\mathbf{h}_q$, from the source to microphone are available, the dereverberation problem can be formulated as an inverse problem in form of:
\begin{equation}\label{eq:inverse}
\min_{\textbf{s}[t]} \sum_{t=1}^{T} \big| {b}_{q}[t] - \textbf{h}_q ^{\top}\textbf{s}[t]\big|^{2} +  \mathcal{J}_{\rm reg}(\{\textbf{s}[t]\}_{t=1}^T).
\end{equation}
with  $\mathcal{J}_{\rm reg}$ {being} a regularizer {which characterizes}  priors of $\textbf{s}[t]$.  Problem formulation~\eqref{eq:inverse}  shares similarities with problems encountered in the image processing community, such as image deconvolution \cite{wang2020learning}, which has been successfully addressed using the Plug-and-Play (PnP) strategy. However, applying an inverse filtering strategy for dereverberation presents challenges due to the estimate of AIRs~\cite{naylor2010speech}. Therefore, most recent work instead resort to the WPE formulation for dereverberation.

\subsection{Vanilla WPE method}\label{sec:wpe}
%Here, we apply the theory of MCLP that has been shown effective for dereverberation, where the observed reverberant speech with a time delay is utilized to predict the late-reverberant components and the prediction residual is considered to be desired speech. For the sake of convenience to formulate the problem, signal model Eq.~\eqref{eq:signal} can be approximated in short-time Fourier transform (STFT) domain by~\cite{nakatani2010speech}

The WPE algorithm belongs to the MCLP class and is typically applied in the frequency domain. The signal model described in Eq.~\eqref{eq:signal} can be approximated in the Short-Time Fourier Transform (STFT) domain~\cite{nakatani2010speech}
\begin{equation}\label{eq:sigstft}
\begin{aligned}
 B_{q}(n,k) = \sum_{l=0}^{{L-1}}H_{q}(l,k) S(n-l,k),
\end{aligned}
\end{equation}
where $n$ and $k$ are the time-frame and frequency bin indices respectively. $B_{q}(n,k)$, $S(n,k)$ and ${H}_{q}(n,k)$ represent the counterparts of ${b}_q[t]$, $s[t]$ and ${h}_{q,l}$ in STFT domain respectively. According to the theory of MCLP, the signal model can be converted to the following auto-regressive problem:
\begin{equation}\label{eq:wpematrix}
\begin{aligned}
B_{\text{ref}}(n,k) = \hat{S}(n,k)+ \overline{\mathbf{w}}^{\rm H}(k)\overline{\mathbf{b}}(n-D,k),
\end{aligned}
\end{equation}
%\begin{equation}\label{eq:wpesig}
%B_{\text{ref}}(n,k) = \hat{S}(n,k)+ \sum_{l=D}^{D+L-1} {W}(l,k){S}(n-l,k)
%\end{equation}
%\begin{equation}\label{eq:wpesig}
%\begin{aligned}
%&B_{\text{ref}}(n,k) = \hat{S}(n,k)+ \sum_{l=D}^{D+L-1} {W}(l,k){S}(n-l,k)\\
%&=\sum_{l=0}^{D-1} {H}(n,k){S}(n-l,k) + \sum_{l=D}^{D+L-1} {W}(l,k){S}(n-l,k),
%\end{aligned}
%\end{equation}
where $B_{\text{ref}}(n,k)$ is the reference signal randomly {chosen} at any microphone. $\hat{S}(n,k)$ corresponds to the desired signal that consists the direct-path signal and early-reverberation component determined by prediction delay $D$. The regressor $\overline{\mathbf {b}}(n-D,k)$ in frequency bin $k$, of length $L_{Q} = L \times Q$ , is constructed by concatenating the regressors of all $Q$ channels
\begin{equation}\label{eq:b}
   \overline{\mathbf{b}}(n-D,k)= \text{col}\{{\mathbf b}_1(n-D,k), \cdots, {\mathbf b}_Q(n-D,k)\}
 \end{equation}
 with ${\mathbf b}_q(n-D,k) = (B_q(n-D,k),\cdots, B_q(n-D-L+1,k) )^\top$.  $\overline{\mathbf{w}}(k)$ is the prediction filter of length $L_{Q}$ in frequency bin $k$,  constructed by concatenating prediction weights of each channel
\begin{equation}\label{eq:w}
\begin{aligned}
\overline{\mathbf w}(k)= \text{col}\{{\mathbf w}_1(k), \cdots, {\mathbf w}_Q(k)\}
\end{aligned}
\end{equation}
with ${\mathbf w}_q(k) = (W_q(0,k),\cdots, W_q(L-1,k))^\top$.
%and $L$ denotes the order of $H_q(n,k)$
%In addition, each frequency bin of the reverberant signal is modeled independently using a delayed MCLP.
%If modelling each frequency bin of the reverberant signal independently and rewriting Eq.\eqref{eq:wpesig} in vector notation, we can obtain:
%\begin{equation}\label{eq:wpematrix}
%\begin{aligned}
%B_{\text{ref}}(n,k) = \hat{S}(n,k)+ \overline{\mathbf{w}}^{\rm H}(k)\overline{\mathbf{b}}(n-D,k),
%\end{aligned}
%\end{equation}
%where $\overline{\mathbf {b}}(n-D,k)$ is constructed by:
%\begin{equation}\label{eq:b}
%\begin{aligned}
%&\overline{\mathbf{b}}(n-D,k)= \\
% \Big[B_1(n-D,k),&\cdots, B_1(n-D-L+1,k),\cdots\\
% ,B_Q(n-D,k), &\cdots, B_Q(n-D-L+1,k) \Big]^\top,
%\end{aligned}
%\end{equation}
%to form a vector of length $L_{Q} = L \times Q$, and $\overline{\mathbf w}(k)$ is the filter vector constructed by:
%\begin{equation}\label{eq:w}
%\begin{aligned}
%\overline{\mathbf w}(k)&=  \Big[\overline{\mathbf w}_1(0,k),\cdots, \overline{\mathbf w}_1(L-1,k),\cdots,\\
% &\overline{\mathbf w}_{Q}(0,k), \cdots, \overline{\mathbf w}_{Q}(L-1,k) \Big]^\top
%\end{aligned}
%\end{equation}
%of length $L_{Q}$.
With an given $\overline{\mathbf w}(k)$,  the prediction residual is considered as the desired signal that can be evaluated by:
\begin{equation}\label{eq:wpeproblem}
\hat{S}(n,k) = B_{\text{ref}}(n,k) - \overline{\mathbf{w}}^{\rm H}(k)\overline{\mathbf b}(n-D,k).
\end{equation}
%where $\overline{\mathbf w}(k)$ is the inverse filter to be estimated.

The WPE method aims to effectively determine the filter weight vector $\overline{\mathbf w}(k)$. Assuming that the desired signal follows a zero-mean complex Gaussian distribution with a time-varying variance $\sigma(n,k)$, the probability density function can be expressed as:
\begin{equation}\label{eq:distribute}
\begin{aligned}
\mathcal{P}\Big(\hat{S}(n,k)\Big) &= \mathcal{N}_{\mathbb{C}}\Big(\hat{S}(n,k);0,\sigma(n,k)\Big)\\
& = \frac{1}{\pi\sigma(n,k)}e^{-\frac{|\hat{S}(n,k)|^2}{\sigma(n,k)}}
\end{aligned}
\end{equation}
where $\sigma(n,k)$ represents the Power Spectral Density (PSD) of the desired signal and is considered as an unknown parameter. Based on this physical model described in Equation \eqref{eq:distribute}, we can formulate the log-likelihood function for each independent frequency as:
\begin{equation}\label{eq:log}
\begin{aligned}
\mathcal{L}\Big(\overline{\mathbf{w}}(k),\boldsymbol{\sigma}(k)\Big) &= \prod_{n=1}^{N} \mathcal{N}_{\mathbb{C}}\Big(\hat{S}(n,k);0,\sigma(n,k)\Big), \\
 &\quad \text{for all }  k=1,\cdots,K
\end{aligned}
\end{equation}
with ${\boldsymbol{\sigma}}(k) =[\sigma(1,k),\sigma(2,k),\cdots,\sigma(N,k)]^\top$. In order to estimate $\overline{\mathbf w}(k)$ and $\boldsymbol{\sigma}(k)$, a maximum likelihood criterion is employed, leading to minimize the following cost function:
\begin{equation}\label{eq:wpe}
\begin{aligned}
&\mathcal{J}\Big(\{\overline{\mathbf{w}}(k)\}_{k=1}^K,\{\boldsymbol{\sigma}(k)\}_{k=1}^K\Big) \\
&= \sum_{k=1}^K \sum_{n=1}^N \frac{|\hat{S}(n,k)|^2}{\sigma(n,k)}+\log\pi\sigma(n,k).
\end{aligned}
\end{equation}
The estimation of $\overline{\mathbf{w}}(k)$ and $\boldsymbol{\sigma}(k)$ can be performed iteratively. Given $\boldsymbol{\sigma}(k)$, $\overline{\mathbf{w}}(k)$ is obtained using a closed-form solution:
\begin{equation}\label{eq:wpew}
\begin{aligned}
\overline{\mathbf{w}}(k) = [Z_{\overline{\mathbf b}}(k)]^{-1}\mathbf{q}_{\overline{\mathbf b}}(k),
\end{aligned}
\end{equation}
where
\begin{equation}\label{eq:wpeR}
Z_{\overline{\mathbf b}}(k) = \sum_{n=1}^{N}\frac{\overline{\mathbf{b}}(n-D,k)[\overline{\mathbf{b}}(n-D,k)]^{\rm H}}{\sigma(n,k)}
\end{equation}
and
\begin{equation}\label{eq:wpeP}
\mathbf{q}_{\overline{\mathbf b}}(k) = \sum_{n=1}^{N}\frac{\overline{\mathbf{b}}(n-D,k){B_{\text{ref}}^{*}(n,k)}}{\sigma(n,k)}.
\end{equation}
Once $\overline{\mathbf{w}}(k)$ is obtained, we can calculate $\hat{S}(n,k)$ using Eq. \eqref{eq:wpeproblem} and subsequently estimate the PSD of $\hat{S}(n,k)$ as:
\begin{equation}\label{eq:wpesigmaa}
\sigma(n,k) = \text{max}\big\{ |\hat{S}(n,k)|^2, \epsilon \big\}
\end{equation}
with $ \epsilon$, a lower bound needs to be predefined. %The procedure of vanilla WPE is summarized in Algorithm \ref{alg:wpe}.
%\begin{algorithm}[t]
%    \caption{Vanilla WPE for speech dereverberation.}
%    \label{alg:wpe}
%    %\LinesNumbered
%    \KwIn {The observed multi-channel speech:~$\overline{\mathbf{b}}(n-D,k)$}
%    \KwOut { The estimated speech:~$\mathbf{\hat{S}}$}
%    \textbf{Parameters:}~The lower bound:~$\epsilon$; The time delay: D; The filter order: L\\
%     \textbf{Initialize:}  \textbf{Initialize} The predicted filter weight vector:~$\overline{\mathbf{w}}(k)={\underbrace{[0,\cdots,0]}_{L\times Q}}^T$\\
%    \While{ not converged}{
%     \For{k = 1 to K}{
%     \For{n = 1 to N}{
%      Calculate $\sigma(n,k)$ by \eqref{eq:wpesigmaa}\;}
%      Calculate  $Z_{\overline{\mathbf b}}(k)$ by \eqref{eq:wpeR}\;
%      Calculate  $\mathbf{q}_{\overline{\mathbf b}}(k)$ by \eqref{eq:wpeP}\;
%      Update $\overline{\mathbf{w}}(k)^{(l+1)}$ by \eqref{eq:wpew}\;}
%    }
%Conduct inverse filtering by \eqref{eq:wpeproblem}
%\end{algorithm}
\section{Plug-and-play framework for speech dereverberation}
In this section, we first reformulate the prediction error minimization problem of WPE to incorporate deep speech priors, and then present the solving method.
%\subsection{Signal model}
%We suppose the signal model under the scenario where a distant microphone array with $Q$ channels captures the convolved speech with ambient noise.
\subsection{Problem formulation}
In this work, we consider a multiple-input-single-output~(MISO) scenario, that is, to calculate the filter weight through all received reverberant signals and to estimate the desired signal at a reference microphone that is randomly chose. Meanwhile, the received signal contains both reverberation and noise. Therefore, we assume the scenario where a $Q$-channel array captures the convolved speech with explicit ambient noise and thus the signal model should be defined by
\begin{equation}\label{eq:signalpnp}
\begin{aligned}
x_{q}[t] &= {b}_q[t] + {y}_q[t]\\
&=h_{q}[t]\ast s[t]+ {y}_{q}[t],
\end{aligned}
\end{equation}
where ${y}_{q}[t]$ represents the additive noise independent of ${b}_q[t]$. Likewise, in  STFT domain, \eqref{eq:signalpnp} can be represented as
\begin{equation}\label{eq:sigstftpnp}
\begin{aligned}
X_{q}(n,k) &=  B_{q}(n,k) + {Y}_{q}(n,k) \\
 &= \sum_{l=0}^{{L-1}}H_{q}(l,k) S(n-l,k)+ {Y}_{q}(n,k),
\end{aligned}
\end{equation}
with $Y_{q}(n,k)$ being the STFT representation of $y_{q}[t]$. Considering the MCLP process, the desired speech should be estimated by
\begin{equation}\label{eq:desiredR}
R(n,k) \!=\! X_{\rm ref}(n,k)-\overline{\mathbf{w}}^{\rm H}(k)   \overline{\mathbf{x}}(n-D,k) - V(n,k),
\end{equation}
%where $X_{\rm ref}(n,k)$ corresponding the reference speech of our signal model in Sec.~\ref{sec:sig},
where $V(n,k)$ is an estimate of the noise, and $\overline{\mathbf {x}}(n-D,k)$ is constructed by the way similar to $\overline{\mathbf {b}}(n-D,k)$ in Sec.~\ref{sec:wpe}. The rationale of introducing $V(n,k)$ will be elaborated in the next subsection and confirmed in the experiment. For notation simplicity, we also denote the prediction error without removing noise by:
\begin{equation}\label{eq:desired}
\hat{S}(n,k) \!=\! X_{\rm ref}(n,k)-\overline{\mathbf{w}}^{\rm H}(k)   \overline{\mathbf{x}}(n-D,k).
\end{equation}

From Eq.~\eqref{eq:wpesigmaa}, we can see that the {time-varying} variance $\sigma(n,k)$ is related to the estimated signal and can be evaluated and inserted over iterations. Therefore, we denote the cost function related to WPE by:
\begin{equation}\label{eq:wpenow}
\mathcal{J}_{\rm WPE}\Big(\{\overline{\mathbf{w}}(k)\}_{k=1}^K\Big)\!=\!\sum_{k=1}^K \sum_{n=1}^N \frac{|{\hat{S}}(n,k)|^2}{\sigma(n,k)},
\end{equation}
Since the prediction error $R(n,k)$  is considered as the desired signal, it is beneficial to introduce a regularization term to incorporate {speech priors (with respect to $R(n,k)$) for \eqref{eq:wpenow}}:
\begin{equation}
\label{eq:wpe_reg}
\mathcal{J}_{\rm WPE\_Reg}\Big(\{\overline{\mathbf{w}}(k)\}_{k=1}^K\Big) =  \mathcal{J}_{\rm WPE}\Big(\{\overline{\mathbf{w}}(k)\}_{k=1}^K\Big) + \beta \mathcal{J}_{\rm Reg}\big({\mathbf{R}}\big),
\end{equation}
where  $\beta$ is a trade-off parameter, $\mathcal{J}_{\rm Reg}$ denotes a regularizer, and ${\mathbf{R}}$ is the  speech time-frequency matrix consisting of {${\{{R}(n,k)\}}_{n=1,k=1}^{N,K}$}\footnote{Note that the similar definition will be used for the other bold capital letters, such as $\hat{\mathbf{S}}$, $\mathbf{V}$ and $\mathbf{P}$.}.

Handcrafting a performant regularizer $\mathcal{J}_{\rm Reg}$ and finding an efficient solving method is not a trivial task. Instead, we propose to learn priors from speech data and incorporate them into the mathematics-based optimization based on the PnP strategy. The prototype PnP scheme, however, presents some complexities. The regularization is implicit as it involves a denoising algorithm, and careful parameter tuning is required during the PnP procedure, as the choice of the denoiser plugged into the framework affects the input noise-level of the denoising algorithm~\cite{romano2017little}. To ensure a more general regularization for our framework, we consider $\mathcal{J}_{\rm Reg}$ in the following form:
\begin{equation}
\label{eq:red}
         \mathcal{J}_{\rm Reg}\big({\mathbf{R}}\big) =  \frac{1}{2}\big<{\mathbf{R}},{\mathbf{R}}-\Omega ({\mathbf{R}})\big>,
\end{equation}
where  $\Omega(\cdot)$ denotes an off-the-shelf denoiser.  This specific form is known as RED~\cite{romano2017little}, which has proven to be an effective regularizer. RED exhibits favorable derivative properties under mild assumptions, enabling it to effectively manage the gradient of the regularization term. Notably, the penalty itself is proportional to the inner product between the desired speech, $\mathbf{R}$, and its denoised residual, $\mathbf{R}-\Omega (\mathbf{R})$. This interpretation aligns with a speech-adaptive Laplacian. Incorporating this form~\eqref{eq:red}, the cost function can be expressed as follows:
\begin{equation}
\begin{aligned}
\label{eq:optired}
        & \mathcal{J}_{\rm WPE\_Reg}\Big(\{\overline{\mathbf{w}}(k)\}_{k=1}^K\Big) \\=& \mathcal{J}_{\rm WPE}\Big(\{\overline{\mathbf{w}}(k)\}_{k=1}^K\Big)+\frac{\beta}{2} {\mathbf{R}}^{\rm H}\big[{\mathbf{R}}-\Omega ({\mathbf{R}})\big].
\end{aligned}
\end{equation}
With the inclusion of this explicit regularization expression, our overall objective function becomes clear, well-defined, and more precise. This formulation offers flexibility in accommodating arbitrary denoising engines represented by $\Omega(\cdot)$. %Moreover, the regularization term in our approach is \cred{convex}, providing a guarantee for the convergence of our proposed algorithm.

\subsection{Solving with the variable splitting techqniue}

To solve the problem in \eqref{eq:optired}, we employ the augmented Lagrange technique on $R(n,k)$. This allows us to formulate the problem with additional equality constraints, resulting in the following full problem formulation:
\begin{equation}\label{eq:J3}
\begin{aligned}
 \min_{\overline{\mathbf{w}}(k),\mathbf{R},\mathbf{V}}  &\sum_{k=1}^K\sum_{n=1}^N\frac{|{\hat{S}}(n,k)|^2}{\sigma(n,k)}+\frac{\beta}{2} \mathbf{R}^{\rm H}\big[\mathbf{R}-\Omega(\mathbf{R})\big]\\
\text{s.t.} \quad R(n,k) &= X_{\rm ref}(n,k)-\overline{\mathbf{w}}^{\rm H} (k)  \overline{\mathbf{x}}(n-D,k) - V(n,k) \\
% \quad\mathcal{E}(\mathbf{R}) &= \sigma_{\rm norm}^2, \\
 ~\text{for}~ n &= 1,\cdots,N~\text{and}~k =1,\cdots,K,
\end{aligned}
\end{equation}
%where $V(n,k)~\text{with}~n=1,\cdots, N~\text{and}~k = 1, \cdots, K$~denote the auxiliary variables which differ from the typical PnP procedure. For simplicity, we denote these auxiliary variables as $\mathbf{V}$.
%which will be explained subsequently in detail.
%$\mathcal{E}(\cdot)$ represents the energy of the argument signal, and $ \sigma_{\rm norm}^2$ denotes a predefined signal energy.

%Note that there is one part in problem~\eqref{eq:J3} that , i.e., we introduce auxiliary variables $V(n,k)$ with $n=1,\cdots,N ~\text{and}~k=1,\cdots,K$ which can be denoted as $\mathbf{V}$, which is differ

%We have two points of view to elaborate on the rationale behind the introduction of the term $\mathbf{V}$. From the perspective of mathematics, introducing new variables $\mathbf{V}$ is able to avoid the overdetermination of the first constraint in \eqref{eq:J3}. Specifically, if we rewrite the this constraint without such auxiliary term for each frequency, we will obtain:
We have two perspectives to further explain the rationale behind introducing the term $V(n,k)$.
\begin{itemize}
\item From a mathematical standpoint, the introduction of new variables $V(n,k)$ helps to prevent the overdetermination issue in \eqref{eq:J3}. To be specific,  if we were to rewrite the constraints without the auxiliary term for each frequency,  we have $N \times K$ equality constraints, while the number of unknowns is $L_Q \times K$. The vanilla WPE problem is inherently overdetermined, which limits the effectiveness of adding extra regularization. This is unlike regular PnP based on \eqref{eq:inverse}, which is usually underdetermined in many applications. Therefore, we add $\mathbf{V}$ with $N\times K$ variables so that the regularization has sufficient degree of freedom in the solution space.

  \item Furthermore, from a physical modeling perspective, we interpret $\mathbf{R}$ as the clean output speech without any reverberation or noise. The vanilla WPE is specifically designed for speech dereverberation and is theoretically not suitable for speech denoising. To address this, we explicitly introduce the  additive noise term during problem modeling and processing. By doing so, we aim to account for the presence of noise and maintain the physical interpretation in the formulation.
 \end{itemize}
% Under such assumption, considering the situation without noise, we can set $\mathbf{V}$ to 0 and simplify the optimization process, then $\mathbf{R}$  reduces to $\hat{\mathbf{S}}$ in~\eqref{eq:wpe_reg} and~\eqref{eq:red}.

%With respect to the reason why we introduce the energy constraint to~\eqref{eq:J3} is that we need it to ensure the uniqueness of the solution. To be more specific, during the optimization iterations, $\mathbf{V}$ could be extraordinary large to output extremely small value of $\mathbf{R}$, so as to further minimize problem~\eqref{eq:J3}. Intuitively, the inherent structural characteristics of speech does not dependent on the specific energy range, and thus the small value of $\mathbf{R}$ still plays a role in incorporating speech prior to our problem, whereas it could be deleterious to intelligibility for humankind with such small energy of speech. Therefore, we conduct energy normalization for $\mathbf{R}$ to restrict the energy within a reasonable range, which is to avoid indeterminacy to problem~\eqref{eq:J3} caused by the new term $\mathbf{V}$ to some extent.

In order to facilitate the solving process, we convert the constrained problem to the unconstrained problem in \eqref{eq:J3} by the corresponding (scaled) augmented Lagrangian function, which is defined as:
\begin{equation}\label{eq:L}
\begin{aligned}
&\mathscr{L}\Big(\overline{\mathbf{w}}(k)_{k=1}^K, \mathbf{R}, \mathbf{V}, \mathbf{P}\Big)\\ =& \mathcal{J}_{\rm WPE}+\frac{\beta}{2} \mathbf{R}^{\rm H}\big[\mathbf{R}-\Omega(\mathbf{R})\big] +\frac{\rho}{2}\sum_{k=1}^K\sum_{n=1}^N\Big(\big|[X_{{\rm ref}}(n,k)-\overline{\mathbf{w}}^{\rm H}(k) \\ &\times\overline{\mathbf{x}}(n-D,k)]\!-\!V(n,k)\!-\!R(n,k)\!+\!P(n,k)\big|^2\!-\!|P(n,k)|^2\Big),
\end{aligned}
\end{equation}
where $P(n,k)$ is the scaled dual variable, and $\rho$ is the penalty parameter. We then iteratively optimize the primal and dual variables of \eqref{eq:L} by solving subproblems over iteration index ${\ell}$ as follows.

\begin{enumerate}[]
\item Step 1 --- Optimization with respect to $\overline{\mathbf{w}}(k)$: The optimization of \eqref{eq:L} reduces to
\begin{equation}\label{eq:w}
\begin{aligned}
\overline{\mathbf{w}}(k)
&= \mathop{\rm argmin}_{\overline{\mathbf{w}}(k)}\mathcal{J}_{\rm WPE}+\frac{\rho}{2}\sum_{k=1}^K\sum_{n=1}^N\big|X_{\rm ref}(n,k)\\&-[\overline{\mathbf{w}}(k)]^{\rm H}\overline{\mathbf{x}}(n-D,k)-V^{(\ell)}(n,k)\\&-R^{(\ell)}(n,k)+P^{(\ell)}(n,k)\big|^2.
\end{aligned}
\end{equation}
The optimization w.r.t. $\overline{\mathbf{w}}(k)$ is a separable least square problem and can be then solved by
\begin{equation}\label{eq:w1}
\overline{\mathbf{w}}^{(\ell+1)}(k) = [{R_{\overline{\mathbf{x}}}^{(\ell+1)}(k)}]^{-1}\mathbf{p}_{\overline{\mathbf{x}}}^{(\ell+1)}(k),
\end{equation}
where
\begin{equation}\label{eq:R}
{R_{\overline{\mathbf{x}}}^{(\ell+1)}(k)}= \sum_{n=1}^{N}\frac{\overline{\mathbf{x}}(n-D,k)[\overline{\mathbf{x}}(n-D,k)]^{\rm H}}{\lambda^{(\ell+1)}(n,k)}
\end{equation}
and
\begin{equation}\label{eq:pp}
\mathbf{p}_{\overline{\mathbf{x}}}^{(\ell+1)}(k) = \sum_{n=1}^{N}\frac{\overline{\mathbf{x}}(n-D,k)\tilde{X}^{(\ell+1)}(n,k)^{*}}{\lambda^{(\ell+1)}(n,k)}.
\end{equation}
In the above solution,  $\lambda^{(\ell+1)}(n,k)$ is given by
\begin{equation}\label{eq:lamda}
\lambda^{(\ell+1)}(n,k) = \frac{2\sigma^{(\ell)}(n,k)}{2+\rho\sigma^{(\ell)}(n,k)},
\end{equation}
where $\sigma^{(\ell)}(n,k)$ is calculated by \eqref{eq:wpesigmaa}, and $\tilde{X}^{(\ell+1)}(n,k)$ is given by
\begin{equation}\label{eq:tldx}
\begin{aligned}
  \tilde{X}^{(\ell+1)}(n,k)=& X_{{\rm ref}}(n,k)- \frac{\rho}{2}\lambda^{(\ell+1)}(n,k)  \big[R^{(\ell)}(n,k)\\ & \quad+V^{(\ell)}(n,k)-P^{(\ell)}(n,k)\big].
\end{aligned}
\end{equation}
By substituting $\overline{\mathbf{w}}(k)$ of each band into \eqref{eq:desired}, we can construct matrix $\hat{\mathbf{S}}$ which will be used in the following steps.

\item Step 2 --- Optimization with respect to $\mathbf{R}$: The optimization problem (\ref{eq:L}) now reduces to
\begin{equation}\label{eq:r}
\begin{aligned}
\mathbf{R}^{(\ell+1)} =& \mathop{\rm argmin}_{\mathbf{R}}\frac{\rho}{2}\big\|\hat{\mathbf{S}}^{(\ell+1)}-\mathbf{V}^{(\ell)}-\mathbf{R}^{(\ell)}+\mathbf{P}^{(\ell)}\big\|^2
\\+&\frac{\beta}{2} [\mathbf{R}^{(\ell)}]^{\rm H}\big[\mathbf{R}^{(\ell)}-\Omega(\mathbf{R}^{(\ell)})\big].
\end{aligned}
\end{equation}
From the perspective of RED \cite{romano2017little}, the prior properties of speech can be incorporated in \eqref{eq:r} by applying a denoising processing to speech $\tilde{\mathbf{R}}$ defined by
\begin{equation}\label{eq:tildeR}
\tilde{\mathbf{R}}^{(\ell+1)}=\hat{\mathbf{S}}^{(\ell+1)}-\mathbf{V}^{(\ell)}+\mathbf{P}^{(\ell)}.
\end{equation}
%Substituting it to \eqref{eq:r}, we can obtain
%\begin{equation}\label{eq:tildeR}
%\begin{aligned}
% \mathbf{R}^{(\ell+1)} &= \mathop{\rm argmin}_{\mathbf{R}}\frac{\rho}{2}\big\|\mathbf{R}^{(\ell)}-\tilde{\mathbf{R}}^{(\ell+1)}\big\|^2\\&+\frac{\beta}{2} [\mathbf{R}^{(\ell)}]^{\rm H}\big[\mathbf{R}^{(\ell)}-\Omega(\mathbf{R}^{(\ell)})\big].
%\end{aligned}
%\end{equation}
Setting the derivative of the above optimization w.r.t.~$\mathbf{R}$ to zero, we can minimize it directly by some iterative scheme, leading to the following equation:
\begin{equation}\label{eq:graR}
\rho\big(\mathbf{R}^{(\ell)}-\tilde{\mathbf{R}}^{(\ell+1)}\big)+\beta\big(\mathbf{R}^{(\ell)}-\Omega(\tilde{\mathbf{R}}^{(\ell)})\big)=0.
\end{equation}
The solution to this problem can be achieved via the fixed-point iteration:
\begin{equation}\label{eq:withdenoiser}
\begin{aligned}
&\rho\big(\mathbf{R}^{(\ell,i+1)}-\tilde{\mathbf{R}}^{(\ell+1)}\big)+\beta\big(\mathbf{R}^{(\ell,i+1)}-\Omega(\tilde{\mathbf{R}}^{(\ell,i)}\big)=0\\
\end{aligned}
\end{equation}
leading to
\begin{equation} \label{eq:withdenoiser2}
\mathbf{R}^{(\ell+1,i+1)} = \mu \tilde{\mathbf{R}}^{(\ell+1)}+(1-\mu) \Omega(\tilde{\mathbf{R}}^{(\ell,i)};\Theta)
\end{equation}
with $\mu = \frac{\rho}{\rho+\beta}$ and inner iteration $i = 1,\cdots,I$, where $\Theta$ denotes the parameters of the denoiser.
%Meanwhile, considering the energy normalization in~\eqref{eq:J3}, we conduct the normalization
%\begin{equation}\label{eq:r4}
%\mathbf{R}^{(\ell+1,i+1)} = \frac{\mathbf{R}^{(\ell+1,i+1)}}{\mathcal{E}(\mathbf{R}^{(\ell+1,i+1)})}\sigma_{\rm norm}^2,
%\end{equation}
%where  $\sigma_{\rm norm}^2 = \mathcal{E}(\mathbf{\hat{S}}^{(\ell+1)})$.
\item Step 3 --- Optimization with respect to $\mathbf{V}$: Here, the solution of this optimization problem readily writes
\begin{equation}\label{eq:epsilon}
\mathbf{V}^{(\ell+1)} = \hat{\mathbf{S}}^{(\ell+1)}-\mathbf{R}^{(\ell+1)}+\mathbf{P}^{(\ell)}.
\end{equation}
\item Step 4 --- Update of $\mathbf{P}$:  This dual variable is updated in the standard manner:
\begin{equation}\label{eq:p}
\mathbf{P}^{(\ell+1)}= \mathbf{P}^{(\ell)}+ \hat{\mathbf{S}}^{(\ell+1)}
-\mathbf{V}^{(\ell+1)}-\mathbf{R}^{(\ell+1)}.
\end{equation}
\end{enumerate}
Variables $\overline{\mathbf{w}}(k)$, $\mathbf{R}$, $\mathbf{V}$ and $\mathbf{P}$ are updated until convergence, and output $\mathbf{R}$ will be used as the estimated speech.
\begin{algorithm}[t]
    \caption{Plug-and-play framework for speech dereverberation.}
    \label{alg:pnp}
    \KwIn {The observed multi-channel speech:~$\overline{\mathbf{x}}(n-D,k)$}
    \KwOut { The estimated speech:~$\mathbf{R}$}
    \textbf{Parameters:}~The time delay: D; The filter order: L; The lower bound:~$\epsilon$; The inner iterations:~I; The penalty parameter of ADMM:~$\rho$; The parameter of the fixed-point algorithm: $\mu$\\
     \textbf{Initialize:}   The auxiliary variables:~$\mathbf{V} = \boldsymbol{0}_{N\times K} $; The dual variables:~$\mathbf{P} =  \boldsymbol{0}_{N\times K} $; The predicted filter weight vector:~$\overline{\mathbf{w}}(k)= \boldsymbol{0}_{L_Q\times 1} $.\\
    \While{ not converged}{
    \For{k = 1 to K}{
     \For{n = 1 to N}{
      Calculate $\sigma^{(\ell+1)}(n,k)$ by \eqref{eq:wpesigmaa}\;
      Calculate $\lambda^{(\ell+1)}(n,k)$ by \eqref{eq:lamda}\;
      Calculate $\tilde{X}^{(\ell+1)}(n,k)$ by \eqref{eq:tldx};}
      Calculate  ${R_{\overline{\mathbf{x}}}^{(\ell+1)}(k)}$ by \eqref{eq:R}\;
      Calculate  $\mathbf{p}_{\overline{\mathbf{x}}}^{(\ell+1)}(k)$ by \eqref{eq:pp}\;
      Update $\overline{\mathbf{w}}(k)^{(\ell+1)}$ by \eqref{eq:w1}\;}
      Update ${\mathbf{\hat{S}}^{(\ell+1)}}$  by \eqref{eq:desired}\;
      Update $\tilde{\mathbf{R}}^{(\ell+1)}$ by \eqref{eq:tildeR}\;
      \For{i = 1 to I}{
        Update $\mathbf{R}^{(\ell+1,i)}$ by \eqref{eq:withdenoiser2} (denoising) \;
%        Normalize $\mathbf{R}^{(l+1,i)}$ by \eqref{eq:r4};
        }
     Update $\mathbf{V}^{(\ell+1)}$ by \eqref{eq:epsilon}\;
     Update $\mathbf{P}^{(\ell+1)}$ by \eqref{eq:p}  \;
     $\ell = \ell+1$ }
\end{algorithm}
\setlength{\textfloatsep}{0pt}

\begin{figure*}[!t]
 \centering
      \includegraphics[width=1.0\textwidth]{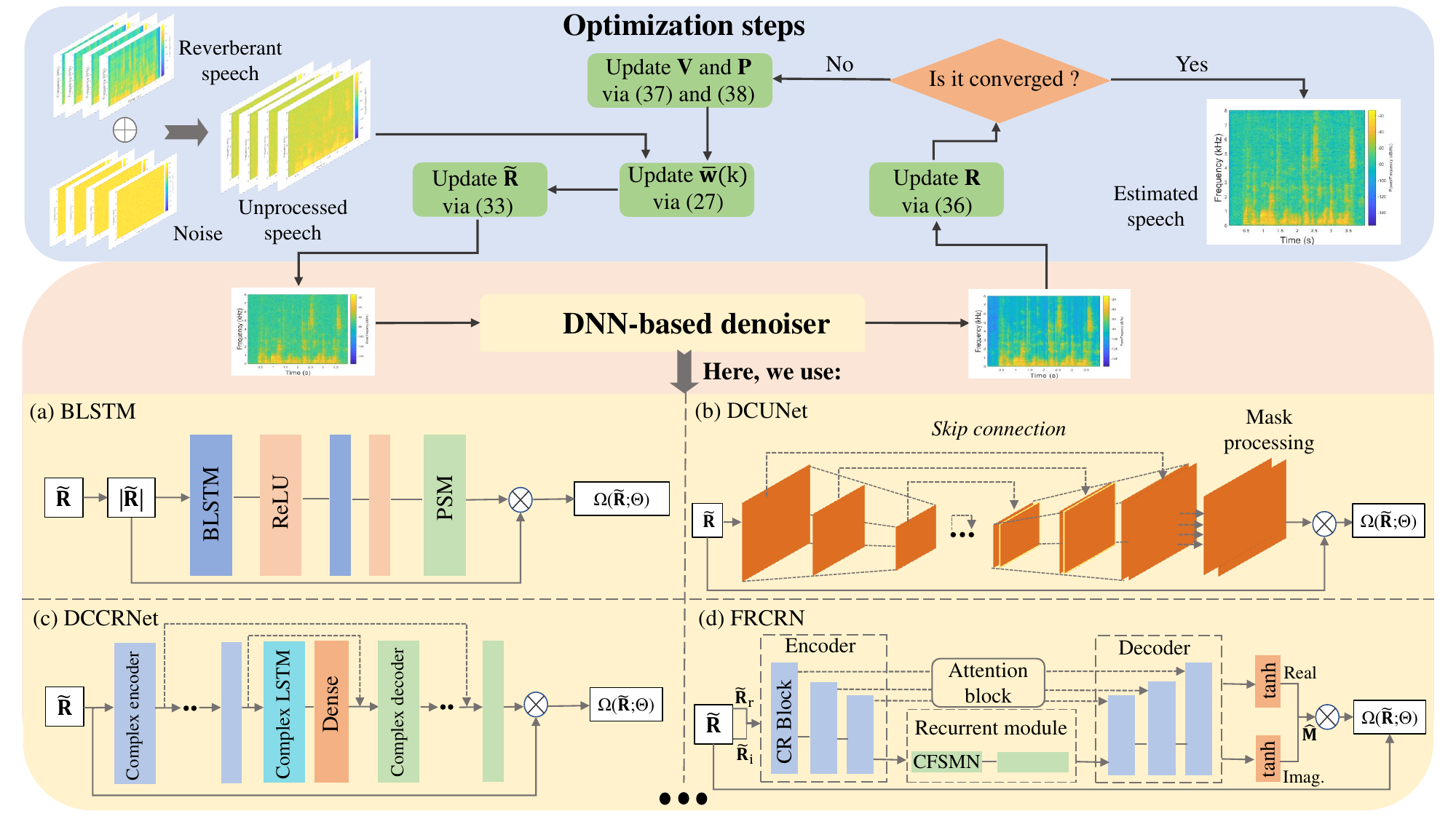}
       \vspace{-5mm}
  \caption{Diagram of PnPWPE. The upper part illustrates the optimization steps of our proposed framework, and the bottom part shows the way we insert the DNN-based denoiser into PnPWPE as well as  network structures of denoisers we utilized in this work.}
  \label{fig:nn}
   \vspace{-3mm}
\end{figure*}
\section{Experiments}
In this section, we first present the DNN-based denoisers inserted in PnPWPE, then introduce the comparison methods and evaluation metrics utilized in our experiments, and finally validate the effectiveness of PnPWPE in several respects.
%\begin{table*}[t!]\footnotesize
%\renewcommand\arraystretch{1.4}
%\caption{The comparison results of WPE and the proposed method under the scenario of reverberation without noise.}\label{table:t2}\vskip 5pt
%\centering
%\begin{tabular}{c|c|c|c|c|c}
%\toprule
%Compared methods &Dereverberation & Denoise &Domain &Training dataset&Model\\\cline{1-6}
%	Conv-TasNet	& \Checkmark& \Checkmark&Time	&WHAMR~\cite{maciejewski2020whamr}&The fully-convolutional TasNet~\cite{luo2019conv}\\\cline{1-6}
%	WPE	&\Checkmark &\XSolidBrush&Time-frequency&	\textbf{-}&	\textbf{-}\\\cline{1-6}
%BLSTM	&\XSolidBrush&\Checkmark&Time-frequency	&WSJ0Mix&	BLSTM\\\cline{1-6}
%	DCUNet&\XSolidBrush	&\Checkmark&Complex&	 LibriMix~\cite{cosentino2020librimix}&	Deep Complex U-net~(Large-DCUnet-20)~\cite{choi2019phase}\\\cline{1-6}
%	DCCRN	&\XSolidBrush&\Checkmark&Complex&	LibriMix&	Deep Complex CRN~\cite{hu2020dccrn}\\\cline{1-6}
%	FRCRN&&\Checkmark&	Complex&	DNS-challenge~\cite{dubey2022icassp}&	Frequency Recurrent CRN~\cite{zhao2022frcrn}\\
%\bottomrule
%%&Conv-TasNet&WPE&BLSTM&DCUNet&DCCRNet&FRCRN\\\cline{1-7}
%%		
%%	&Time&Time-frequency	&Time-frequency&Complex	&Complex	&Complex\\\cline{1-7}
%%&WHAMR&\textbf{-}&WSJ0Mix&LibriMix&LibriMix&DNS-challenge\\	\cline{1-7}
%%&The fully-convolutional TasNet&\textbf{-}&BLSTM&Deep Complex U-net&Deep Complex CRN&Frequency Recurrent CRN\\\cline{1-7}
%\end{tabular}
%\end{table*}
\begin{table*}[t!]\footnotesize
\renewcommand\arraystretch{1.4}
\caption{Details about DNN-based denoisers which are plugged into the proposed framework.}\label{table:nn}
\centering
\begin{tabular}{c|c|c|c|c|c}
\toprule
 DNN-based denoiser&Dereverberation & Denoise &Domain &Training dataset&Neural network\\\cline{1-6}
	%Conv-TasNet	& \Checkmark& \Checkmark&Time	&WHAMR~\cite{maciejewski2020whamr}&The fully-convolutional TasNet~\cite{luo2019conv}\\\cline{1-6}
%	WPE	&\Checkmark &\XSolidBrush&Time-frequency&	\textbf{-}&	\textbf{-}\\\cline{1-6}
BLSTM	&\XSolidBrush&\Checkmark&Time-frequency	&WSJ0Mix& Bidirectional Long Short-term Memory Network\\\cline{1-6}
	DCUNet~\cite{choi2019phase}&\XSolidBrush	&\Checkmark&Complex&	 Libri1Mix~\cite{cosentino2020librimix}&Deep Complex U-net~(Large-DCUnet-20)\\\cline{1-6}
	DCCRN~\cite{hu2020dccrn}	&\XSolidBrush&\Checkmark&Complex&	Libri1Mix&Deep Complex CRN\\\cline{1-6}
	FRCRN~\cite{zhao2022frcrn}&\Checkmark&\Checkmark&	Complex&	DNS-Challenge~\cite{dubey2022icassp}&Frequency Recurrent CRN\\
\bottomrule
%&Conv-TasNet&WPE&BLSTM&DCUNet&DCCRNet&FRCRN\\\cline{1-7}
%		
%	&Time&Time-frequency	&Time-frequency&Complex	&Complex	&Complex\\\cline{1-7}
%&WHAMR&\textbf{-}&WSJ0Mix&LibriMix&LibriMix&DNS-challenge\\	\cline{1-7}
%&The fully-convolutional TasNet&\textbf{-}&BLSTM&Deep Complex U-net&Deep Complex CRN&Frequency Recurrent CRN\\\cline{1-7}
\end{tabular}
\vspace{-3mm}
\end{table*}
\subsection{Data-driven prior construction}\label{sec:data-driven}

Benefiting from the generalization ability of the proposed framework, any denoiser can be plugged into it to incorporate speech priors. In order to illustrate such flexilty of our framework, in this work, we apply 4 different kinds of DNN-based denoisers trained by different datasets. The network structure diagrams are depicted in Fig.~\ref{fig:nn} and training details are summarized in Table~\ref{table:nn}. In particular, we initially train a time-frequency (T-F) domain-based denoiser for illustrative purposes. Subsequently, we integrate three pre-trained and open-source {complex domain denoisers} into our framework. This approach showcases the versatility of our framework in accommodating various denoisers and leveraging their capabilities to enhance speech quality.
\subsubsection{BLSTM-based denoiser}
%To focus on the main idea of this work,  here we simply train a {bidirectional long short-term memory (BLSTM)}-based denoiser for illustrative purpose.
As illustrated in Fig.~\ref{fig:nn}(a), we train a bidirectional long short-term memory (BLSTM)-based denoiser
%containing two combined layers from bottom-up followed by a mask estimation layer. Each combined layer consists of a BLSTM layer and a Rectifier Linear Unit (ReLU) activation function layer.
%Applying magnitude of the noisy speech (denoted by $|\mathbf{{Z}}|$) as the input feature, the network is trained
to predict the phase sensitive mask (PSM, \cite{wang2018supervised}) for target speech via the magnitude and temporal spectrum approximation loss, defined by~\cite{xu2019optimization}
\begin{equation}
%\begin{split}
\begin{aligned}
\label{eq:cost}
&\mathcal{J}_{\text{Denoiser}}\!=\!\frac{1}{N}\sum \Big(\parallel\mathbf{\hat{M}}\odot |\mathbf{{Z}}|-|\mathbf{A}|\odot \cos(\theta_{\mathbf{{Z}}} -\theta_{\mathbf{A}})\parallel^2_F\\
   &+\!w_d\parallel f_d(\mathbf{\hat{M}}\odot |\mathbf{{Z}}|)-f_d(|\mathbf{A}|\odot \cos(\theta_{\mathbf{{Z}}} -\theta_{\mathbf{A}}))\parallel^2_F\\
   &+\!w_c\parallel f_c(\mathbf{\hat{M}}\odot |\mathbf{{Z}}|)-f_c(|\mathbf{A}|\odot \cos(\theta_{\mathbf{{Z}}} -\theta_{\mathbf{A}}))\parallel^2_F\Big),
\end{aligned}
\end{equation}
where $\hat{\mathbf M}$ denotes the predicted PSM,  $\odot$ is the Hadamard product, $|\mathbf{Z}|$and $|\mathbf{A}|$ are the magnitude of noisy and clean speech, and $\theta_{\mathbf{{Z}}}$ and $\theta_{\mathbf{A}}$ represent phase angles of noisy speech and the clean speech respectively.
% \cred{PSM i.e.,~$\mathbf{M}$, is given by}
%\begin{equation}
%{\mathbf{M}} = \frac{|{\mathbf{A}|}}{|\mathbf{Z}|}\odot\cos(\theta_{\mathbf{{Z}}} -\theta_{\mathbf{{A}}}).
%\end{equation}

 Taking the dynamic information into consideration, we employ the functions (i.e., $f_d(\cdot)$ and $f_c(\cdot)$) to calculate increment and acceleration,
 %\cite{furui1986speaker},
 the increment computation function is defined by
 \begin{equation}
f_d\big(c[n]\big)= \frac{\sum_{o=1}^{O}o\times \big(c[n+o]-c[n-o]\big)}{\sum_{o=1}^{O}2o^{2}},
\end{equation}
where $c[n]$ denotes the magnitude of a time frame, and  $O$ represents the contextual window which is set to 2. The acceleration function can be calculated by computing the increment twice. The weights $w_d$ and $w_c$ are set to 4.5 and 10.0 respectively.

To train a blind BLSTM-denoiser, we add the diffuse noise set, containing white Gaussian noise~(WGN), babble noise, cafe noise, factory noise and bus noise chosen from NOISE-92 database~\cite{varga1993assessment} and chime-3 dataset~\cite{barker2015third}, to the clean speech signals which are randomly chosen from the Wall Street Journal dataset~(WSJ0,~\cite{garofolo1993csr}). Then, we obtain a dataset termed WSJ0Mix, where the training set contains 20,000 utterances and the validation set contains 5000 utterances at various signal-noise ratios (SNRs) between -5 dB and 40 dB. Each utterance in WSJ0Mix is split into 4 seconds with a sampling rate of 16 kHz. We implement the BLSTM-based denoiser based on Adam optimizer \cite{kingma2014adam} with  an initial learning rate {of} 0.0005 and a mini-batch of 32 to minimize the loss function (\ref{eq:cost}) in 60 epochs.
\subsubsection{DCUNet-based denoiser}
Deep Complex U-Net~(DCUNet) is trained to estimate complex ratio mask for target speech based on a variant of U-Net~\cite{ronneberger2015u}. As illustrated in Fig.~\ref{fig:nn}(b), DCUNet consists of a complex convolutional autoencoder with skip-connections. In our work, we insert larger DCUnet-20~(Large-DCUnet-20)\footnote{The open-source DCUNet is available at \url{https://huggingface.co/JorisCos/DCUNet_Libri1Mix_enhsingle_16k}.} trained by libri1mix dataset into our framework since the authors in \cite{choi2019phase} claim that this network with more channels in each layer outperforms other architectures.

\subsubsection{DCCRNet-based denoiser}
Deep Complex Convolution Recurrent Network~(DCCRNet)~\cite{hu2020dccrn} conducts complex-valued speech enhancement via the convolutional encoder-decoder~(CED) architecture with a complex LSTM layer between the encoder and the decoder, as shown in Fig.~\ref{fig:nn}(c). DCCRNet not only benefits from 2D convolution layer (Conv2d) to extract high-level features for better enhancement but also benefits from LSTM to model temporal context with less parameters. Here, we plug DCCRN-CL\footnote{The open-source DCCRNet is available at \url{https://huggingface.co/JorisCos/DCCRNet_Libri1Mix_enhsingle_16k}.}which is also trained by Libri1Mix dataset into our framework.
\subsubsection{FRCRN-based denoiser}
Frequency Recurrent Convolutional Recurrent Networks~(FRCRN), as shown in Fig.~\ref{fig:nn}(d), is trained to predict complex ideal ratio mask
%\cite{williamson2015complex}
for target speech via an extended CED architecture, known as convolutional recurrent encoder-decoder structure which is able to improve feature representation. In addition, the recurrent module and attention block of FRCRN are designed to model the long-term temporal dependencies and to facilitate information flow respectively. The FRCRN model\footnote{The open-source FRCRN is available at \url{https://www.modelscope.cn/models/damo/speech_frcrn_ans_cirm_16k}.} we insert into our framework is trained by Deep Noise Suppression (DNS)-Challenge dataset~\cite{dubey2022icassp},  of which 30\% of dataset contains reverberant components.

Once the denoiser has been already pre-trained, it can be directly plugged into the proposed framework, yielding Algorithm \ref{alg:pnp}.
\subsection{Method comparison and evaluation}

\subsubsection{Comparison method}
Our objective is to enhance the performance of the vanilla WPE algorithm in complex environments. Therefore, in our experiments, we consider the vanilla WPE as the baseline for comparison.

Since our experimental scenarios involve both additive noise and reverberation, we also explore a straightforward concatenation of denoising and dereverberation algorithms to ensure a fair comparison. We introduce two  comparison methods, denoted by \texttt{Denoise+WPE} and \texttt{WPE+denoise}, with different orderings of the algorithms. In the \texttt{Denoise+WPE} approach, we initially apply the denoising algorithm to each channel of the unprocessed speech signals. The enhanced signals from all channels are then concatenated along the channel axis to obtain a multi-channel signal, which is further processed by the vanilla WPE algorithm. On the other hand, for \texttt{WPE+denoiser}, considering the MISO (Multiple-Input Single-Output) characteristic of the WPE method, we apply the single-channel dereverberated signal to the denoising algorithm after applying the vanilla WPE. To ensure a fair comparison, the denoising algorithms used in these approaches are the same ones we integrate into our proposed framework.

Apart from simple combination of denoising and dereverberant algorithms, we compare our proposed framework with convolutional time-domain audio separation network~(Conv-TasNet)~\cite{luo2019conv}, an end-to-end time domain speech separation algorithm. We utilize an open-source Conv-TasNet model\footnote{The open-source Conv-TasNet is available at \url{https://huggingface.co/cankeles/ConvTasNet_WHAMR_enhsingle_16k}.} to directly process the test set. Since this model has been pre-trained by the single-speaker version of WHAMR!~\cite{maciejewski2020whamr} with corpus containing noise and reverberation, it is considered as a pure dataneural-network~(NN) comparison method in our experiments.
\subsubsection{Evaluation metrics}

For evaluation metrics, we adopt three widely used measures for speech dereverberation, i.e., perceptual evaluation of speech quality (PESQ) \cite{rix2001perceptual}, cepstral distance (CD) \cite{kinoshita2016summary} and frequency-weighted segmental signal-to-noise ratio~(F-SNR)~\cite{kinoshita2016summary} in our experiments. In general, for PESQ and F-SNR, larger values indicate better performance, while for CD, smaller values indicate better performance. To explicitly show the convergence of proposed framework, we define energy error between the left and right sides of the constraint of \eqref{eq:J3}:
\begin{equation}
\label{eq:t60}
\text{Error} = \frac{1}{NK}\sum_{n=1}^{N}\sum_{k=1}^{K}|R(n,k)-\hat{S}(n,k)- V(n,k)|^2.
\end{equation}
%where $N$ and $K$ are the number of frames and frequency bands.

\subsection{Results}
\subsubsection{Dataset description}
\begin{table}[t!]\footnotesize
\renewcommand\arraystretch{1.2}
\caption{Parameter setting of generating test sets.}\label{table:data}
\centering
\begin{tabular}{c|c|c}
\toprule
Room type &Room A&Room B\\ \cline{1-3}
Room length (m)&8-13&15-20\\ \cline{1-3}
Room width (m)&8-13&15-20\\ \cline{1-3}
Room height (m)&2.8-3.8&3-4\\ \cline{1-3}
T60 (ms) &400-800&800-1200\\ \cline{1-3}
Minimum distance &\multirow{2}{*}{0.5}&\multirow{2}{*}{0.8}\\
from source to wall (m) &&\\ \cline{1-3}
Minimum distance &\multirow{2}{*}{0.8}&\multirow{2}{*}{1.3}\\
from source to microphone (m)&&\\ \cline{1-3}
\multirow{3}{*}{Microphone array} & \multicolumn{2}{c}{Linear array with 4 channels}\\ \cline{2-3}
 & \multicolumn{2}{c}{Inner distance = 40 mm}\\ \cline{2-3}
  & \multicolumn{2}{c}{Microphone type = Omnidirectional}\\
  \bottomrule
\end{tabular}
\end{table}

To test the proposed framework, we generate a test set by simulating random reverberant scenarios. Specifically, we randomly generate 4 microphones in two types of rooms, denoted as Room A and Room B, corresponding to different room size ranges. The microphone array and speech source are located randomly within the rooms. More details can be found in Table~\ref{table:data}. We create room impulse responses (RIRs) using the method in~\cite{allen1979image}. We split 20 clean speech signals, randomly chosen from the Librispeech corpus~\cite{panayotov2015librispeech}, into 4-second segments with a sampling rate of 16 kHz. We then convolve them with the RIRs, where the reverberant time (T$_{60}$) is constrained within the range of 400-800 milliseconds (ms) for Room A and 800-1200 ms for Room B.

In addition, we evaluate the robustness of the proposed framework under noisy and reverberant conditions. As widely demonstrated in image processing based on PnP, this strategy to excavate deep priors is able to yield promising results with White Gaussian Noise (WGN). However, for speech signal processing, it is insufficient to only test the performance in the case of WGN, and thus we consider more scenarios. In our experiments, we choose three additive noises, namely WGN, babble, and cafe, chosen from the diffuse noise set (mentioned in Sec.~\ref{sec:data-driven}), and further add them to the reverberant speech signals. We set the Signal-to-Noise Ratios (SNRs) to 0 dB, 10 dB, and 20 dB, respectively. This way, we obtain a test set of 18 experimental scenarios and a total of 360 utterances. To avoid the randomness of the results, we evaluate the methods on the test set and take the average value.

\subsubsection{Experimental settings}

\begin{figure}[t!]
\centering
       \includegraphics[width=0.5\textwidth]{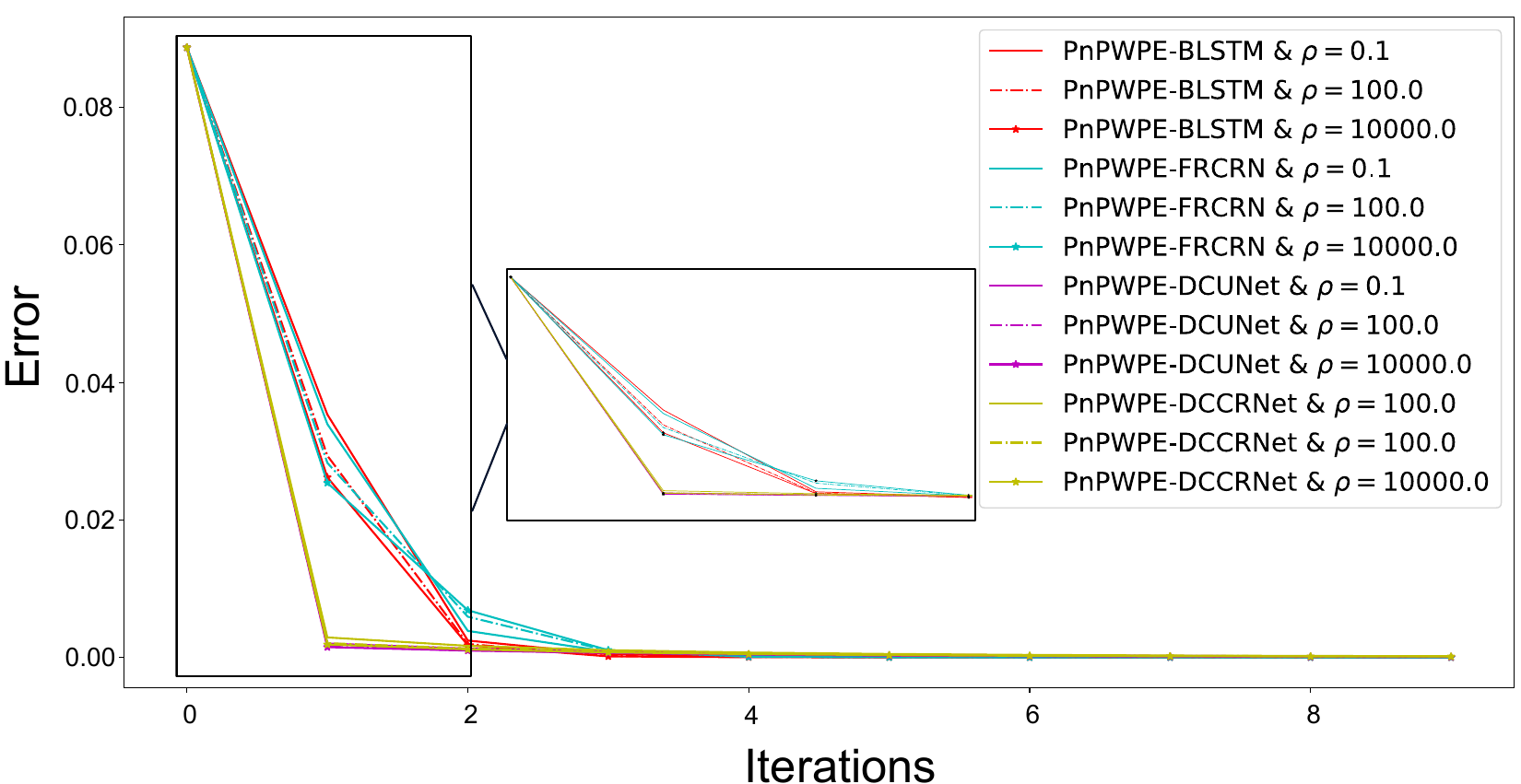}
        \vspace{-5mm}
  \caption{The Error convergence curves of the proposed framework in terms of different settings of $\rho$ in the scenario with WGN at SNR = 0 dB in Room B. }
  \label{fig:rho}
   \vspace{-3mm}
\end{figure}

\begin{figure}[t!]
\centering
       \includegraphics[width=0.5\textwidth]{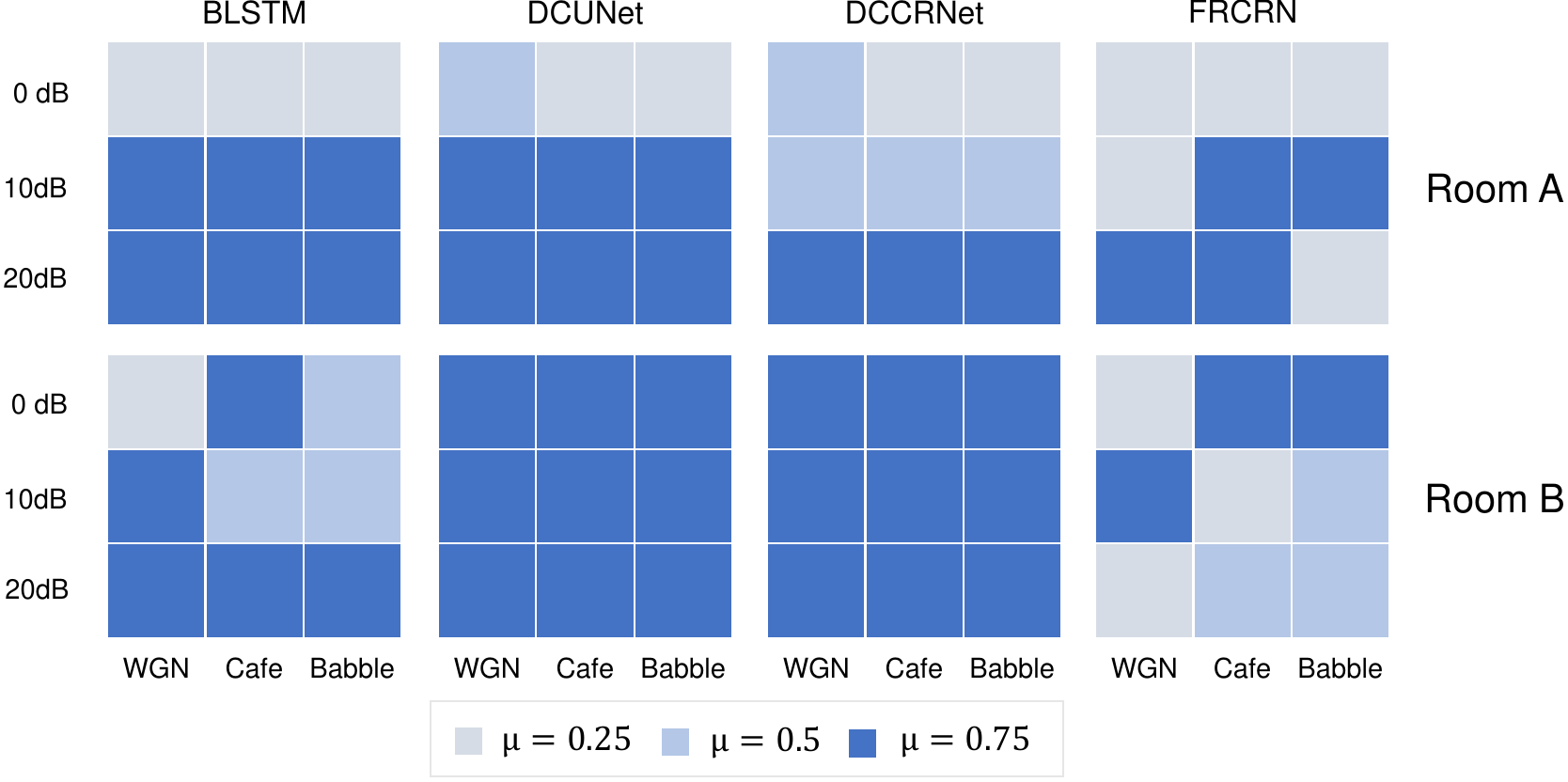}
        \vspace{-5mm}
  \caption{The settings of $\mu$ in different scenarios, where each 3$\times$3 square denotes PnPWPE inserted with a kind of DNN-based denoiser.}
  \label{fig:mu}
\end{figure}

The proposed framework is implemented in the STFT domain using a Hann window, where the frame length is 32 ms of 75\% overlapping. For MCLP process, considering the trailing effect of reverberation, the filter order $L$ is
%calculated by:
%\begin{equation}
%\label{eq:t60}
%L = \frac{\rm{T}_{60}\times\rm{sample\_rate}}{\rm{frame\_length}}.
%\end{equation}
%Therefore, by substituting the maximum $T_{60}$ of each room into $\eqref{eq:t60}$,$L$ is
set to 28 for Room A and 35 for Room B. In addition, we set the time delay $D$ to 2 and choose the first microphone as the reference microphone in all experiments. For calculating $\sigma$ in  \eqref{eq:wpesigmaa}, the lower bound~$\epsilon$ is set to 0.0001. %In the test, we find that the proposed framework would suffer from heavy loads and time consuming because of the update of $\overline{\mathbf w}(k)$ and several activations of denoiser during the fixed-point iteration~\cite{romano2017little}.
For simplicty, we set the inner iteration $I$ to 1 to speed the overall framework.  The penalty parameter $\rho$ in the augmented Lagrangian function only affects the convergence speed, as theoretically shown~\cite{zhao2021plug} and experimentally shown in Fig.~\ref{fig:rho}. Therefore we simply set $\rho = 0.1$ in our experiments. As for the trade-off parameter $\mu$ in the fixed-point iteration controlling the impact of the denoising engine, we first choose a relatively suitable setting on a single signal for each experimental scenario, and then apply the selected setting to all test sets. The specific settings of $\mu$ are summarize in Fig.~\ref{fig:mu}.

\subsubsection{Results discussion}
\begin{table*}[t!]\footnotesize
\renewcommand\arraystretch{1.2}
\caption{The results of all comparison methods in Room A with all types of noises of different levels in Room A. The best results are in bold and the second best results are underlined.}\label{table:r1}
\centering
%\renewcommand{\arraystretch}{<1.3>}
%\begin{tabular}{|c|c|c|c|c|c|c|c|c|c|c|c|}
\begin{tabular}{c|c|c|ccc|ccc|ccc}
\toprule
\multirow{2}{*}{Additive noise type}&\multicolumn{2}{c|}{Noise level} &\multicolumn{3}{c|}{SNR = 0 dB} &\multicolumn{3}{c|}{SNR = 10 dB} &\multicolumn{3}{c}{SNR = 20 dB}\\ \cline{2-12}
%&Method & Model &PESQ $\uparrow$ &CD $\downarrow$&fwSNRseg$\uparrow$ &PESQ$\uparrow$ &CD $\downarrow$ &fwSNRseg$\uparrow$ &PESQ$\uparrow$ &CD $\downarrow$&fwSNRseg$\uparrow$\\ \cline{1-12}
&Method & Denoiser &PESQ  &CD &F-SNR &PESQ &CD &F-SNR &PESQ &CD &F-SNR\\ \cline{1-12}
\midrule
\multirow{16}{*}{WGN}&Unprocessed & \textbf{-} & 1.420 &8.474 &4.720 &1.933 &7.974  & 5.724& 2.076&6.692 &6.311 \\ \cline{2-12}
&WPE & \textbf{-} & 1.313 & 8.628&4.338 &1.870 & 8.351 & 5.735& 2.281&7.265& 7.366\\ \cline{2-12}
&Conv-TasNet & \textbf{-} & 1.353&9.043 &4.784 & 1.917&  8.255& 5.837 & 2.015&7.182 &5.938 \\ \cline{2-12}
&\multirow{4}{*}{Denoise $+$ WPE} & BLSTM & 1.563 &8.626 &4.638 &1.957 &8.275  &5.646 & 2.538& 7.173& 7.074\\ \cline{3-12}
& & DCUNet & 1.284 & 8.522& 4.666&1.653 &8.074  & 5.740& 2.025&6.765 &6.559 \\ \cline{3-12}
& & DCCRNet & 1.165 &8.505 & 4.693& 1.932& 8.044 & 5.701&1.906 &6.766 &6.455 \\ \cline{3-12}
& & FRCRN & \underline{1.948} &8.550 &4.558 &2.221 & 8.237 &5.689 &2.453 &6.998 &6.480 \\ \cline{2-12}
&\multirow{4}{*}{WPE $+$ Denoise} & BLSTM &1.486  & 7.938& 3.998& 1.832& 7.358 &5.156 & 2.240&6.441 &5.985 \\ \cline{3-12}
& & DCUNet &  1.467& 8.877&4.917 & 2.062&  8.099& 6.181&2.586 &7.303 &6.495 \\ \cline{3-12}
& & DCCRNet & 1.104 & 8.678&4.225 &1.861 & 9.357 &5.731 & 2.296&8.327 &6.079 \\ \cline{3-12}
& & FRCRN & 1.936 & 6.816& 5.873& 2.261& 6.651 &6.157 & 2.651&\underline{6.154} &6.524 \\ \cline{2-12}
&\cellcolor{gray!30}  & \cellcolor{gray!30}BLSTM& \cellcolor{gray!30} 1.695&\cellcolor{gray!30}\underline{6.789} &\cellcolor{gray!30}\underline{6.353} &\cellcolor{gray!30}\underline{2.287} &\cellcolor{gray!30} \underline{6.812} & \cellcolor{gray!30}\textbf{7.397}&\cellcolor{gray!30} \underline{2.659}& \cellcolor{gray!30}\textbf{5.328}& \cellcolor{gray!30}{7.837}\\ \cline{4-12}
& \cellcolor{gray!30}&\cellcolor{gray!30} DCUNet & \cellcolor{gray!30}1.407 & \cellcolor{gray!30}8.444&\cellcolor{gray!30}5.122 & \cellcolor{gray!30}2.175& \cellcolor{gray!30}7.705 &\cellcolor{gray!30}6.746 & \cellcolor{gray!30}2.642& \cellcolor{gray!30}6.239&\cellcolor{gray!30}7.678 \\ \cline{3-12}
& \cellcolor{gray!30}& \cellcolor{gray!30}DCCRNet & \cellcolor{gray!30} 1.202& \cellcolor{gray!30}8.889& \cellcolor{gray!30}4.785&\cellcolor{gray!30}2.034 & \cellcolor{gray!30}8.363 &\cellcolor{gray!30}6.472 &\cellcolor{gray!30}2.535 &\cellcolor{gray!30} 6.656&\cellcolor{gray!30} 7.651\\ \cline{3-12}
& \cellcolor{gray!30}\multirow{-4}{*}{PnPWPE}& \cellcolor{gray!30}FRCRN &\cellcolor{gray!30}\textbf{1.972}  &\cellcolor{gray!30} \textbf{6.342}&\cellcolor{gray!30} \textbf{6.378}&\cellcolor{gray!30} \textbf{2.444}& \cellcolor{gray!30}\textbf{5.687} &\cellcolor{gray!30}\underline{7.145} &\cellcolor{gray!30}\textbf{2.672} &\cellcolor{gray!30}\textbf{5.328}&\cellcolor{gray!30}\underline{7.838} \\ \cline{1-12}

\midrule
\multirow{16}{*}{Cafe}&Unprocessed & \textbf{-} & 1.312 & 6.919&3.036 &2.035 & 6.194 &4.596 &2.083 &5.473 &6.066 \\ \cline{2-12}
&WPE & \textbf{-} & 1.629 & 7.045& 2.934&2.028 & 6.186 & 4.685&2.321 &4.746 &6.733 \\ \cline{2-12}
&Conv-TasNet & \textbf{-} & 1.527 & 6.905&4.544 &1.981 & 6.135 &5.558 & 2.232&5.430 &6.503 \\ \cline{2-12}
&\multirow{4}{*}{Denoise $+$ WPE} & BLSTM & 1.042 & 6.868& 2.847& 1.714& 6.245 &4.352 &2.334& 5.190&6.143 \\ \cline{3-12}
& & DCUNet & 1.354 &6.938 & 2.983& 2.019& 6.209 &4.465 &2.139 &5.267 &6.100 \\ \cline{3-12}
& & DCCRNet & 1.321 & 6.917&3.015 &1.870 & 6.205 & 4.437&2.001 & 5.394& 6.013\\ \cline{3-12}
& & FRCRN & 1.699 & 7.027& 3.095& 2.064&6.318  & 4.548& 2.334& 5.516& 6.101\\ \cline{2-12}
&\multirow{4}{*}{WPE $+$ Denoise} & BLSTM & 0.965 & 8.195& 3.290& 1.690& 7.174 &5.035 &2.317& 5.137&6.827 \\ \cline{3-12}
& & DCUNet &1.669  &7.822 & 4.113&2.129 & 7.310 &5.481 & 2.481&6.238 &7.227 \\ \cline{3-12}
& & DCCRNet &1.579  & 8.489&\textbf{ 4.887}& 1.673& 7.201 & 5.918&2.313 & 7.258&6.790 \\ \cline{3-12}
& & FRCRN &  1.704& \underline{6.444}& \underline{4.816}& 2.197&5.914  &{5.999} & \underline{2.577}&4.722 &\underline{7.463} \\ \cline{2-12}
&\cellcolor{gray!30}& \cellcolor{gray!30}BLSTM & \cellcolor{gray!30}1.541 &\cellcolor{gray!30} 6.791&\cellcolor{gray!30}3.623 &\cellcolor{gray!30} 2.240& \cellcolor{gray!30}5.834 &\cellcolor{gray!30}5.622 & \cellcolor{gray!30}2.452&\cellcolor{gray!30}4.544 &\cellcolor{gray!30}7.324 \\ \cline{3-12}
&\cellcolor{gray!30} &\cellcolor{gray!30} DCUNet & \cellcolor{gray!30}1.578 &\cellcolor{gray!30}6.711 &4\cellcolor{gray!30}.046 &\cellcolor{gray!30} \underline{2.327}& \cellcolor{gray!30} 5.629& \cellcolor{gray!30}5.775& \cellcolor{gray!30} 2.545&\cellcolor{gray!30}\underline{4.450} &\cellcolor{gray!30}7.213 \\ \cline{3-12}
& \cellcolor{gray!30}& \cellcolor{gray!30}DCCRNet & \cellcolor{gray!30}\underline{1.824} &\cellcolor{gray!30}6.850 &\cellcolor{gray!30}4.790 & \cellcolor{gray!30}2.254&  \cellcolor{gray!30} \underline{5.524}&\cellcolor{gray!30}\underline{6.149} & \cellcolor{gray!30}2.477 &\cellcolor{gray!30}4.504& \cellcolor{gray!30}7.251\\ \cline{3-12}
& \cellcolor{gray!30}\multirow{-4}{*}{PnPWPE} &\cellcolor{gray!30} FRCRN &\cellcolor{gray!30} \textbf{1.880} & \cellcolor{gray!30}\textbf{6.335}& \cellcolor{gray!30}4.406& \cellcolor{gray!30}\textbf{2.335}& \cellcolor{gray!30}\textbf{5.519} &\cellcolor{gray!30}5.733 & \cellcolor{gray!30}\textbf{2.579}&\cellcolor{gray!30} \textbf{4.334}&\cellcolor{gray!30}{7.342} \\ \cline{1-12}

\midrule

\multirow{16}{*}{Babble}&Unprocessed & \textbf{-} & 1.324 &6.815 &2.952 &1.935 &6.095  & 4.479& 2.066&5.612 & 5.599\\ \cline{2-12}
&WPE & \textbf{-} & 1.317 & 7.011&2.620 &2.014 & 5.818 &4.311 & 2.421&4.676 &6.216 \\ \cline{2-12}
&Conv-TasNet & \textbf{-} & 1.181 &7.239 &\textbf{4.173} &1.978 & 6.008 &5.577 & 2.062&5.428 &6.077 \\ \cline{2-12}
&\multirow{4}{*}{Denoise $+$ WPE} & BLSTM & 1.221 &6.907 &2.984 &1.654 & 6.097 & 4.139&2.110 &5.416 &5.503 \\ \cline{3-12}
& & DCUNet &  1.061&6.865 & 2.813& 1.476& 5.975 &4.245 & 2.106&5.325 & 5.581\\ \cline{3-12}
& & DCCRNet & 1.138 & 6.856& 2.801&1.977 &  6.012&4.227 & 2.080&5.459 &5.556 \\ \cline{3-12}
& & FRCRN &1.571  &6.894 &2.766 &2.134 &6.156  & 4.308 &2.480 & 5.690& 5.510\\ \cline{2-12}
&\multirow{4}{*}{WPE $+$ Denoise} & BLSTM & 0.989 & 7.946&3.674 &1.659 & 6.741 &5.269 & 2.453& 5.321&7.097 \\ \cline{3-12}
& & DCUNet & 1.092 &8.172 & 3.567& 2.085& 7.020 & 5.497& 2.596&6.043 &7.188 \\ \cline{3-12}
& & DCCRNet & 1.248 &8.627 &3.747 & 1.981&  7.552&5.527 &2.436 & 7.603&6.574 \\ \cline{3-12}
& & FRCRN & 1.307 &6.868 &3.847 & \underline{2.236}&\underline{5.050}  & \textbf{6.269}& \underline{2.626}&\underline{4.037} &\underline{7.629} \\ \cline{2-12}
& \cellcolor{gray!30} & \cellcolor{gray!30}BLSTM & \cellcolor{gray!30}1.523 & \cellcolor{gray!30}\underline{6.542}& \cellcolor{gray!30}\underline{4.025}& \cellcolor{gray!30}2.114&\cellcolor{gray!30} 5.283 &\cellcolor{gray!30}5.610 & \cellcolor{gray!30}\textbf{2.682}&\cellcolor{gray!30}4.257 &\cellcolor{gray!30}7.156 \\ \cline{3-12}
&  \cellcolor{gray!30}&\cellcolor{gray!30} DCUNet & \cellcolor{gray!30}1.311 &\cellcolor{gray!30}7.047 &\cellcolor{gray!30} 3.495& \cellcolor{gray!30}2.168& \cellcolor{gray!30}5.649 & \cellcolor{gray!30}5.083& \cellcolor{gray!30}2.611& \cellcolor{gray!30}4.677&\cellcolor{gray!30}6.682 \\ \cline{3-12}
&  \cellcolor{gray!30}&\cellcolor{gray!30} DCCRNet & \cellcolor{gray!30}\textbf{1.626} &\cellcolor{gray!30} 7.076& \cellcolor{gray!30}3.941&\cellcolor{gray!30} 2.205& \cellcolor{gray!30}5.681 & \cellcolor{gray!30}5.477&\cellcolor{gray!30}2.589 & \cellcolor{gray!30}4.504&\cellcolor{gray!30}7.251 \\ \cline{3-12}
&  \cellcolor{gray!30}\multirow{-4}{*}{PnPWPE}& \cellcolor{gray!30}FRCRN & \cellcolor{gray!30} \underline{1.621}&\cellcolor{gray!30} \textbf{6.528}& \cellcolor{gray!30}3.597&\cellcolor{gray!30} \textbf{2.274}& \cellcolor{gray!30}\textbf{4.883} &\cellcolor{gray!30}\underline{6.216} &\cellcolor{gray!30}2.554 &\cellcolor{gray!30}\textbf{4.025} &\cellcolor{gray!30}{7.550} \\ \cline{1-12}
\bottomrule
\end{tabular}
\vspace{-5mm}
\end{table*}

\begin{figure*}[t!]
\centering
       \includegraphics[width=1\textwidth]{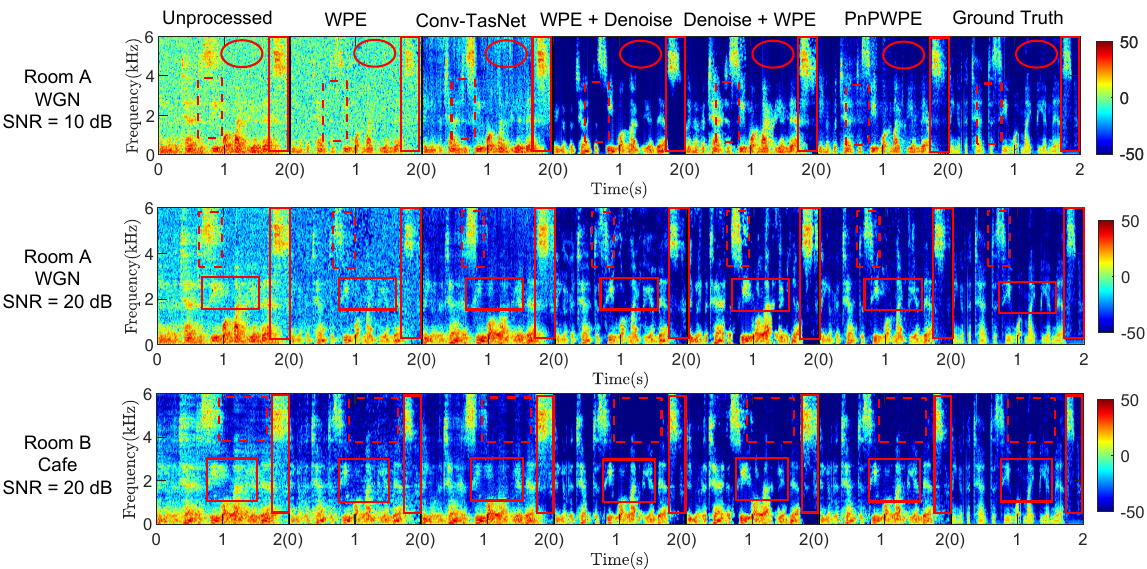}
  \caption{{The visualization results of all comparison methods with the speech truncated to 2 s. From left to right, we present the spectrograms of unprocessed speech, and the estimated speech processed by the vanilla WPE, Conv-TasNet, WPE + denoise, Denoise + WPE, PnPWPE and ground truth respectively.}}
  \label{fig:spec_all}
\end{figure*}

\begin{table*}[t!]\footnotesize
\renewcommand\arraystretch{1.2}
\caption{The results of all comparison methods with all types of noises of different levels in Room B. The best results are in bold and the second best results are underlined.}\label{table:r2}
\centering
%\begin{tabular}{|c|c|c|c|c|c|c|c|c|c|c|c|}
\begin{tabular}{c|c|c|ccc|ccc|ccc}
\toprule
\multirow{2}{*}{Additive noise type}&\multicolumn{2}{c|}{Noise level} &\multicolumn{3}{c|}{SNR = 0 dB} &\multicolumn{3}{c|}{SNR = 10 dB} &\multicolumn{3}{c}{SNR = 20 dB}\\ \cline{2-12}
%&Method & Model &PESQ $\uparrow$ &CD $\downarrow$&fwSNRseg$\uparrow$ &PESQ$\uparrow$ &CD $\downarrow$ &fwSNRseg$\uparrow$ &PESQ$\uparrow$ &CD $\downarrow$&fwSNRseg$\uparrow$\\ \cline{1-12}
&Method & Denoiser &PESQ  &CD &F-SNR &PESQ &CD &F-SNR &PESQ &CD &F-SNR\\ \cline{1-12}
\midrule
\multirow{16}{*}{WGN}&Unprocessed & \textbf{-} & 1.206 & 8.423&4.611 &1.482 &7.935  &5.610 &1.714 &6.749 &6.089 \\ \cline{2-12}
&WPE & \textbf{-} & 1.217 & 8.589& 4.320&1.687 & 8.285 &5.729 & 2.135&7.244 &7.373 \\ \cline{2-12}
&Conv-TasNet & \textbf{-} & 0.667 &9.109 &4.377 & 1.635&8.255  &5.699 & 1.805&7.288 &5.696 \\ \cline{2-12}
&\multirow{4}{*}{Denoise $+$ WPE} & BLSTM & 1.470 &8.702 & 4.574& 1.771& 8.320 & 5.500&2.214 &7.240 & 6.970\\ \cline{3-12}
& & DCUNet & 0.937 &8.467 &4.565 & 1.499& 8.045 &5.522 &1.614 &6.780 & 6.251\\ \cline{3-12}
& & DCCRNet & 0.754 &8.470 &4.593 &1.225 & 8.016 & 5.530&1.437 &6.815 &6.094 \\ \cline{3-12}
& & FRCRN &  1.773& 8.507& 4.469& 1.419& 8.151 &5.447 & 2.169& 7.126& 5.788\\ \cline{2-12}
&\multirow{4}{*}{WPE $+$ Denoise} & BLSTM & 1.461 & 8.001& 3.677& 1.606& 7.471 &4.978 & 1.978&6.506& 5.976\\ \cline{3-12}
& & DCUNet &1.155 & 8.892& 4.569 & 1.805& 8.147 & 6.110& 2.236&7.537 & 6.453\\ \cline{3-12}
& & DCCRNet & 0.714 &9.677 & 3.928& 1.546&9.413  & 5.476& 2.209&8.394 &6.051 \\ \cline{3-12}
& & FRCRN &\underline{1.778}  &\underline{6.915} &5.463 &\underline{2.010} &6.662  &5.971 &\underline{2.387} & 6.112& 6.566\\ \cline{2-12}
&\cellcolor{gray!30}& \cellcolor{gray!30}BLSTM & \cellcolor{gray!30}1.581 &\cellcolor{gray!30}7.017 &\cellcolor{gray!30}\textbf{5.992}&\cellcolor{gray!30}1.981 & \cellcolor{gray!30}\textbf{6.382} &\cellcolor{gray!30}\underline{7.023} & \cellcolor{gray!30}2.276&\cellcolor{gray!30}\underline{5.360} &\cellcolor{gray!30}\underline{7.843}\\ \cline{3-12}
&\cellcolor{gray!30} &\cellcolor{gray!30} DCUNet & \cellcolor{gray!30}1.222 &\cellcolor{gray!30}8.382 &\cellcolor{gray!30}4.945 &\cellcolor{gray!30}1.854 & \cellcolor{gray!30} 7.739& \cellcolor{gray!30}6.509& \cellcolor{gray!30} 2.372&\cellcolor{gray!30} 6.432&\cellcolor{gray!30}7.755\\ \cline{3-12}
& \cellcolor{gray!30}& \cellcolor{gray!30}DCCRNet & \cellcolor{gray!30}1.191 &\cellcolor{gray!30}8.693&\cellcolor{gray!30}4.564& \cellcolor{gray!30}1.823&  \cellcolor{gray!30} 8.234 &\cellcolor{gray!30}6.129 & \cellcolor{gray!30}\textbf{2.397} &\cellcolor{gray!30}6.732& \cellcolor{gray!30}{7.809}\\ \cline{3-12}
& \cellcolor{gray!30}\multirow{-4}{*}{PnPWPE} &\cellcolor{gray!30} FRCRN &\cellcolor{gray!30} \textbf{1.821} & \cellcolor{gray!30}\textbf{6.373}& \cellcolor{gray!30}\underline{5.957}& \cellcolor{gray!30}\textbf{2.080}& \cellcolor{gray!30}\underline{6.446} &\cellcolor{gray!30}\textbf{7.109} & \cellcolor{gray!30}2.325&\cellcolor{gray!30} \textbf{5.173}&\cellcolor{gray!30}7.582 \\ \cline{1-12}
\midrule
\multirow{15}{*}{Cafe}&Unprocessed & \textbf{-} & 1.334 & 6.879& 3.493&1.572 & 6.295 &4.457 &1.671 &5.760 &5.460 \\ \cline{2-12}
&WPE & \textbf{-} & 1.383 &7.057 & 3.263& 1.887&6.169  &4.498 & 2.087& 4.991& 6.007\\ \cline{2-12}
&Conv-TasNet & \textbf{-} & 1.191 & 6.781&\textbf{4.963} & 1.810& 6.320 & 5.408& 1.869& 5.758&5.786 \\ \cline{2-12}
&\multirow{4}{*}{Denoise $+$ WPE} & BLSTM &0.962  &6.845 & 3.437& 1.573& 6.424 & 4.039&1.858 & 5.532& 6.392\\ \cline{3-12}
& & DCUNet & 0.937 & 6.849&3.408 & 1.683& 6.306 &4.295 & 1.704& 5.543& 5.450\\ \cline{3-12}
& & DCCRNet & 0.754 & 6.871& 3.393& 1.579& 6.288 &4.230 & 1.650& 5.700&5.266 \\ \cline{3-12}
& & FRCRN & \underline{1.472}&7.006 & 3.488&2.026 & 6.561 &4.292 & 2.135& 5.859&5.374 \\ \cline{2-12}
&\multirow{4}{*}{WPE $+$ Denoise} & BLSTM & 0.470 & 8.238& 3.197& 1.620& 7.284 & 4.803&1.945 & 5.532& 6.392\\ \cline{3-12}
& & DCUNet & 1.170 & 7.795& 4.280& 1.735&7.175  &5.383 & 2.062& 6.759&6.415 \\ \cline{3-12}
& & DCCRNet & 1.027 & 8.507& \underline{4.894}&1.674 &7.685  &5.300 &1.985 &7.336 &6.075 \\ \cline{3-12}
& & FRCRN & 1.138 &\underline{6.522} &4.830 &1.976 &5.993  &\underline{5.645} & \underline{2.163}& 5.225& \textbf{6.625}\\ \cline{2-12}
&\cellcolor{gray!30}& \cellcolor{gray!30}BLSTM & \cellcolor{gray!30} 1.370&\cellcolor{gray!30}6.817 &\cellcolor{gray!30} 3.818&\cellcolor{gray!30} \underline{2.077}& \cellcolor{gray!30}7.284 &\cellcolor{gray!30}4.803 & \cellcolor{gray!30}2.112&\cellcolor{gray!30} {4.881}&\cellcolor{gray!30}6.372\\ \cline{3-12}
&\cellcolor{gray!30} &\cellcolor{gray!30} DCUNet & \cellcolor{gray!30}1.417 &\cellcolor{gray!30}6.793 &\cellcolor{gray!30}3.939 &\cellcolor{gray!30} 2.000& \cellcolor{gray!30} 5.741& \cellcolor{gray!30}5.377& \cellcolor{gray!30}2.141 &\cellcolor{gray!30}4.884 &\cellcolor{gray!30}6.349\\ \cline{3-12}
& \cellcolor{gray!30}& \cellcolor{gray!30}DCCRNet & \cellcolor{gray!30} \textbf{1.497}&\cellcolor{gray!30}6.637&\cellcolor{gray!30}4.500& \cellcolor{gray!30}2.033&  \cellcolor{gray!30} \textbf{5.597}&\cellcolor{gray!30}\textbf{5.690} & \cellcolor{gray!30}2.125 &\cellcolor{gray!30}4.848& \cellcolor{gray!30}6.412\\ \cline{3-12}
& \cellcolor{gray!30}\multirow{-4}{*}{PnPWPE} &\cellcolor{gray!30} FRCRN &\cellcolor{gray!30} 1.405 & \cellcolor{gray!30}\textbf{6.478}& \cellcolor{gray!30}4.225& \cellcolor{gray!30}\textbf{2.083}& \cellcolor{gray!30}\underline{5.657} &\cellcolor{gray!30}5.617 & \cellcolor{gray!30}\textbf{2.261}&\cellcolor{gray!30} \underline{4.789}&\cellcolor{gray!30} \underline{6.515}\\ \cline{1-12}
\midrule
\multirow{15}{*}{Babble}&Unprocessed & \textbf{-} &1.284  & 6.715& 3.117&1.515 &6.360  &4.274 &1.653 &5.757 &5.185 \\ \cline{2-12}
&WPE & \textbf{-} & 1.622 & 6.871&2.801 & 1.802& 6.068 &4.141 &2.100 &4.846 &5.847 \\ \cline{2-12}
&Conv-TasNet & \textbf{-} & 0.732 & 7.372&3.966 & 1.732&6.225  & 5.356& 1.862&5.663 &5.666 \\ \cline{2-12}
&\multirow{4}{*}{Denoise $+$ WPE} & BLSTM &0.940 & 6.767&2.988 & 1.381&8.372  &3.885 &1.674 &5.618 &4.879 \\ \cline{3-12}
& & DCUNet & 0.663 & 6.756& 2.977&1.541 &6.240  & 3.975& 1.657&5.467 & 5.088\\ \cline{3-12}
& & DCCRNet & 0.957& 6.752& 2.967& 1.653&  6.246& 3.980& 1.709& 5.674&4.897 \\ \cline{3-12}
& & FRCRN & 1.285 & 6.768&2.963 &\textbf{1.977} & 6.457 &4.146 & 2.127& 5.954& 5.407\\ \cline{2-12}
&\multirow{4}{*}{WPE $+$ Denoise} & BLSTM &  0.927& 8.024& 3.264&1.438 &6.919  & 5.165& 2.134& 5.464&6.552 \\ \cline{3-12}
& & DCUNet & 0.840 & 8.321& 3.213&1.631 & 7.203 & 5.234&2.002 & 6.424& 6.780\\ \cline{3-12}
& & DCCRNet &  1.191&8.661 & \underline{3.547}&1.720 & 7.197 & \underline{5.836}&2.054 & 7.594 &6.405  \\ \cline{3-12}
& & FRCRN &  1.001&7.048 & 3.514& 1.866& \underline{5.437} & \textbf{5.941}&2.121 &{4.399} &{6.800}\\ \cline{2-12}
&\cellcolor{gray!30}& \cellcolor{gray!30}BLSTM & \cellcolor{gray!30}1.443 &\cellcolor{gray!30}\textbf{6.628} &\cellcolor{gray!30}\textbf{3.595} &\cellcolor{gray!30}1.936& \cellcolor{gray!30}5.653 &\cellcolor{gray!30}5.209 & \cellcolor{gray!30}2.161&\cellcolor{gray!30} 4.660&\cellcolor{gray!30}6.584\\ \cline{3-12}
&\cellcolor{gray!30} &\cellcolor{gray!30} DCUNet & \cellcolor{gray!30}1.292 &\cellcolor{gray!30}6.808 &\cellcolor{gray!30}3.240 &\cellcolor{gray!30} 1.894& \cellcolor{gray!30} 5.898& \cellcolor{gray!30}4.922& \cellcolor{gray!30}2.154 &\cellcolor{gray!30}4.896 &\cellcolor{gray!30}6.552\\ \cline{3-12}
& \cellcolor{gray!30}& \cellcolor{gray!30}DCCRNet & \cellcolor{gray!30}{1.541} &\cellcolor{gray!30}6.805&\cellcolor{gray!30}3.311& \cellcolor{gray!30}1.913&  \cellcolor{gray!30} 5.872 &\cellcolor{gray!30}5.012 & \cellcolor{gray!30}\underline{2.162} &\cellcolor{gray!30}4.937& \cellcolor{gray!30}6.312\\ \cline{3-12}
& \cellcolor{gray!30}\multirow{-4}{*}{PnPWPE} &\cellcolor{gray!30} FRCRN &\cellcolor{gray!30} \textbf{1.796} & \cellcolor{gray!30}\underline{6.683}& \cellcolor{gray!30}3.267& \cellcolor{gray!30}\underline{1.947}& \cellcolor{gray!30}\textbf{5.355} &\cellcolor{gray!30}5.591 & \cellcolor{gray!30}\textbf{2.239}&\cellcolor{gray!30} \textbf{4.334}&\cellcolor{gray!30}\textbf{6.998} \\ \cline{1-12}
\bottomrule
\end{tabular}
 \vspace{-5mm}
\end{table*}
Table~\ref{table:r1} and Table~\ref{table:r2} summarize the comparison results of all evaluation metrics in Room A and Room B respectively. For visual comparison, we take the scenarios SNR = 10 dB and 20~dB of WGN in Room A and SNR = 20 dB of cafe in Room B as examples, depicted in Fig.~\ref{fig:spec_all}. From overall results, we can briefly draw conclusions on the proposed method as follows:
\begin{itemize}
\item  Our proposed framework is flexible with different denoisers. We observe that for each denoiser, PnPWPE outperforms other combination strategies in most cases. It is worth noting that the DNN-based denoisers we utilized were trained using datasets with various settings. This implies that PnPWPE provides an efficient approach to seamlessly integrate any pre-trained denoiser into the WPE framework in practical applications. Furthermore, we made an interesting discovery that PnPWPE-FRCRN consistently delivers the best results. This can be attributed to the fact that, compared to the other three denoisers, FRCRN was trained using a larger and more complex dataset.%, which also included reverberant data. This finding underscores the importance of a denoiser's ability to accurately model clean speech from reverberant speech, as it greatly enhances the overall effectiveness of the proposed framework.
\item Our proposed framework leverages prior information extracted by DNN-based denoisers, effectively enhancing the quality of reconstructed speech signals in environments characterized by both reverberation and noise. Notably, the performance of PnPWPE-type methods stands out prominently when considering scenarios involving white Gaussian noise (WGN). Table~\ref{table:r1} clearly demonstrates that these methods consistently achieve the best and second-best results. In the case of additive noises in the other two scenarios, while PnPWPE-type methods display some variability in terms of F-SNR values, they still outperform the comparison methods across other evaluation metrics. Moreover, PnPWPE exhibits robustness across different SNRs and reverberation times. Even in challenging conditions characterized by low SNRs and high levels of reverberation, our proposed method maintains its superiority.
\end{itemize}

In what follows, we will first discuss the results between PnPWPE and different types of comparison methods in detail. Then, we will show the usefulness of introducing the term $V(n,k)$, and discuss the convergence property of PnPWPE.

\noindent\textbf{Comparison of PnPWPE and the vanilla WPE:}
 From Table~\ref{table:r1} and Table~\ref{table:r2}, we can see that the proposed PnPWPE shows superiority compared to the vanilla WPE, especially in the scenario with WGN. Specifically, PnPWPE-FRCRN outperforms the vanilla WPE relatively by \textbf{20.7~\% (1.313~$\rightarrow$~1.972)}, \textbf{21.8~\% (1.870~$\rightarrow$~2.444)} and \textbf{17.6~\% (2.281~$\rightarrow$~2.672)}\footnote{The relative improvement is calculated by: $\Delta_{\rm relative} = \frac{\rm PESQ_{\rm 1}-\rm PESQ_{\rm 2}}{4.5-\rm PESQ_{\rm 2}}$ where 4.5 is the upper bound of PESQ.} about PESQ at SNR = 0~dB, 10~dB, 20~dB in Room A,  For Room B, the relative improvements of PESQ are\textbf{ 18.4~\% (1.217~$\rightarrow$~1.821)}, \textbf{14.0~\% (1.687~$\rightarrow$~2.080)} and \textbf{8.0~\% (2.135~$\rightarrow$~2.325)} at SNR = 0~dB, 10~dB, 20~dB respectively. In addition,  Fig.~\ref{fig:spec_all} shows  that the vanilla WPE can not work well under the scenarios with WGN, especially at lower SNRs.

As for more lifelike scenarios, we can observe that, in the cases with cafe and babble which contain the sound of tableware colliding and the chattering of people, the proposed PnPWPE still maintains its advantages compared to the vanilla WPE, as all the evaluation metrics of the former is higher than those of the latter. Actually, it has been found that WPE-type methods are not  robust to additive noise~\cite{huang2022kronecker}. Our results support this view and further demonstrate the necessity of inserting speech prior for WPE to boost the robustness of the vanilla WPE under such complex environments.
%Due to these facts, we can conclude that the proposed framework with prior information excavated by DNN-based denoisers can significantly enhance the reconstructed speech signal under reverberant and noisy environments.

\noindent\textbf{Comparison of PnPWPE and Conv-TasNet:}
From the presented results in Table~\ref{table:r1} and Table~\ref{table:r2}, it can be seen that PnPWPE performs better than Conv-TasNet in most cases. Despite the fact that the F-SNR value of Conv-TasNet is slightly higher than that of PnPWPE-type methods in the scenarios with babble in Room A at SNR = 0~dB, the PESQ and CD value of PnPWPE exceed Conv-TasNet (such as Conv-TasNet V.S. PnPWPE-BLSTM: \textbf{1.181$\rightarrow$1.523, 7.239$\rightarrow$6.542}). The same trend can be found under the cafe noise in Room B at SNR = 0~dB. The main reason why PnPWPE performs more stably than Conv-TasNet may be that the data-driven module in our framework, i.e., the DNN-based denoiser, is to capture speech prior to boost the whole recovery process rather than directly reconstruct the desired speech from the received speech. Therefore, PnPWPE avoids excessive dependence on specific data sets, which is a major factor that leads to poor generalization ability for pure data-driven methods.

From the explicit results in Fig.~\ref{fig:spec_all}, we can find that Conv-TasNet is able to recover clean speech from reverberant and noisy conditions to some extent, but the estimated speech still retains some additive noise that cannot be removed.
Taking the scenarios with WGN as an example, we can clearly see that the estimated speech generated by PnPWPE is more consistent with the ground truth compared to Conv-TasNet.
\begin{figure}[t!]
\centering
       \includegraphics[width=0.5\textwidth]{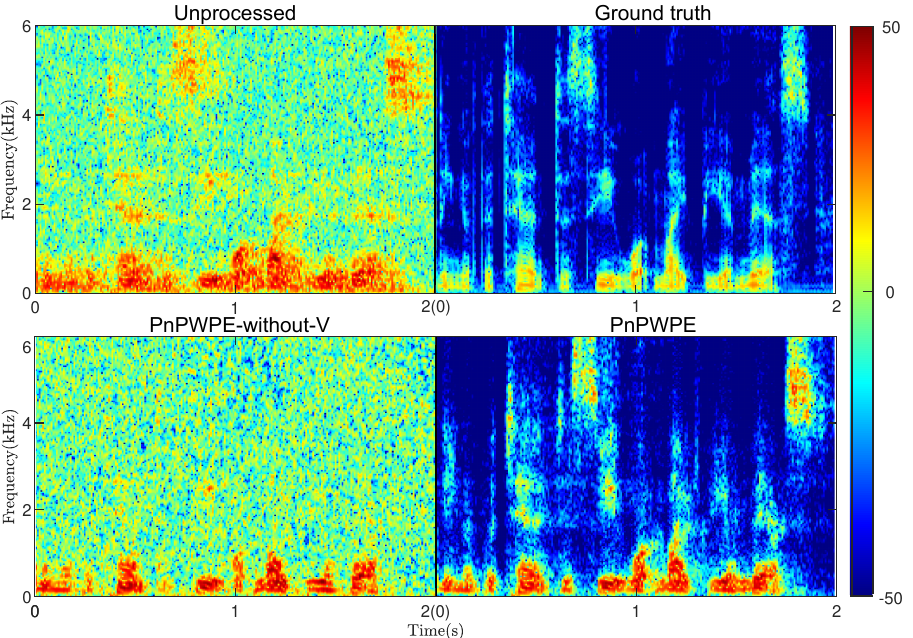}
  \caption{{The visualization results of PnPWPE and PnPWPE-without-$V$ under the scenario with WGN at SNR = 10 dB in Room A.}}
  \label{fig:spec_nov}
\end{figure}
\begin{figure*}[t!]
\centering
       \includegraphics[width=0.32\textwidth]{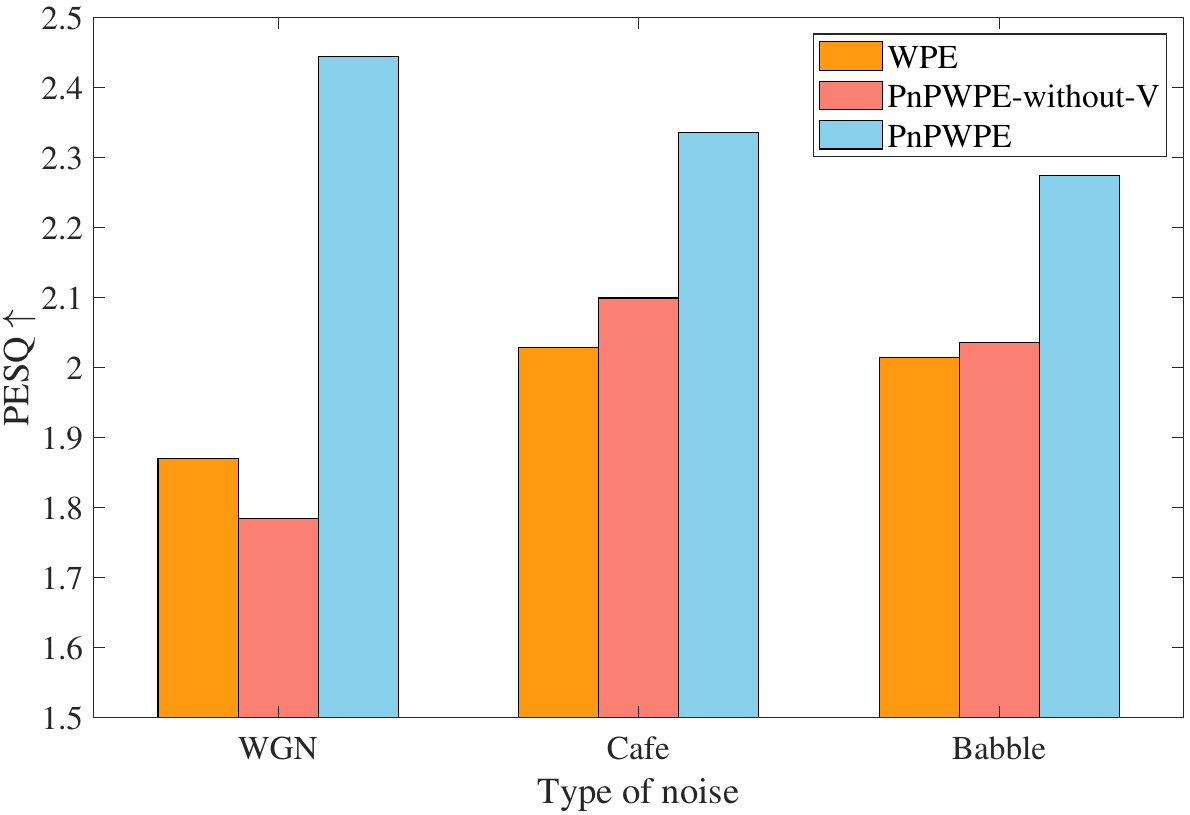}
       \includegraphics[width=0.32\textwidth]{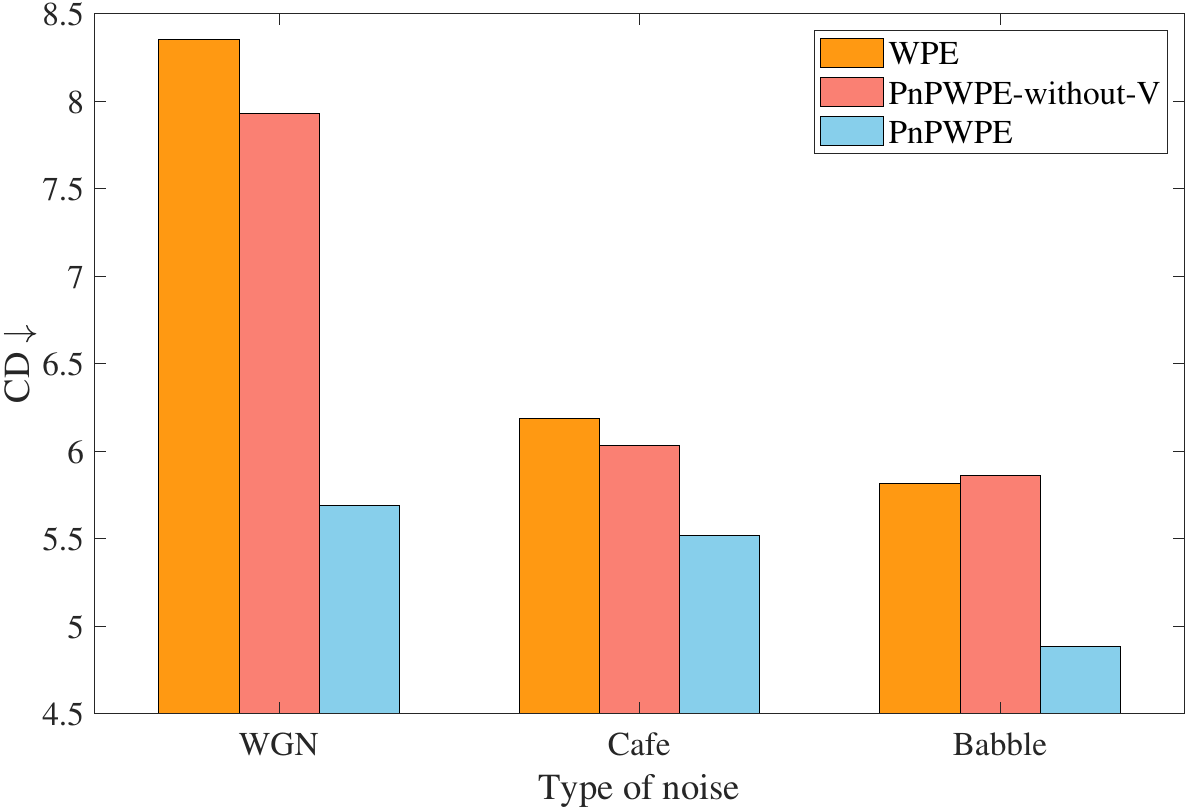}
       \includegraphics[width=0.32\textwidth]{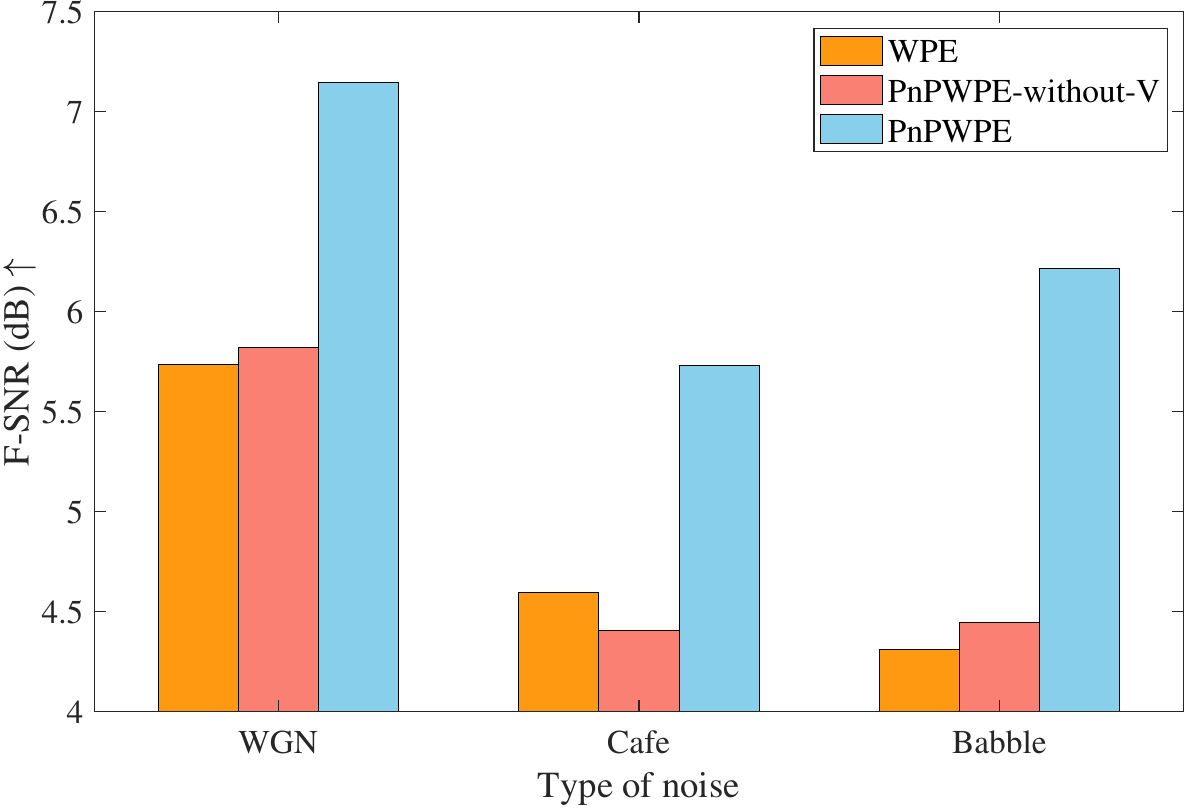}
 \vspace{-3mm}
  \caption{The {bar plot}  of all evaluation metrics in terms of PnPWPE and PnPWPE-without-$V$ under the scenarios with all types of additive noises at SNR=10 dB in Room~A. Here, we insert the FRCRN-based denoiser into the framework as an example.}
  \label{fig:re_nov}
   \vspace{-5mm}
\end{figure*}

\noindent\textbf{Comparison of PnPWPE, Denoise+WPE and WPE+denoise:}
 We compare PnPWPE with different intuitive concatenations of the vanilla WPE and DNN-based denoisers. The experimental results show the superiority of our proposed framework compared to the other concatenated strategies. To be more specific, from Table~\ref{table:r1} and Table~\ref{table:r2}, it can be observed that PnPWPE surpasses \texttt{Denoise+WPE} and \texttt{WPE+denoiser} in most cases, as PnPWPE yields the most best and second best results of PESQ and CD in Room A and Room B. Especially in the scenarios with WGN, the superiority of PnPWPE becomes pronounced as it yields all the best results. Even though in the several conditions with additive noise of people chattering, \texttt{WPE+denoiser} seems to performer better in terms of F-SNR, PnPWPE still maintains its merits if considering PESQ and CD. Taking the scenario of cafe noise (SNR = 0 dB) in Room A as an example, \texttt{WPE+denoiser-DCCRNet} yields the best F-SNR among all the comparison methods. However, the PESQ and CD value of PnPWPE-DCCRNet is \textbf{0.25} and \textbf{1.64} better than the former \textbf{(1.579$\rightarrow$1.824, 8.489$\rightarrow$6.850)}, respectively. The same phenomenon can be found among different denoisers and different scenarios. In addition, we find \texttt{WPE+denoiser} usually performs better than \texttt{Denoise+WPE}, which is probably caused by the fact that denoising first could degrade the spatial property of the observed speech signals, and further impair the performance of  the WPE method.

Examining Fig.~\ref{fig:spec_all}, in the scenario involving white Gaussian noise (WGN) with low signal-to-noise ratio (SNR), the spectrogram of the estimated speech processed by \texttt{WPE+denoiser} reveals the presence of vertical lines (highlighted within an ellipse). These lines correspond to short and impulsive noises. Additionally, as indicated by the dotted rectangles in the figures, the subsequent denoising algorithm of \texttt{WPE+denoiser} tends to excessively cancel the additive noise, particularly in the high-frequency bands. This can result in the loss of desired spectral components, ultimately impacting the quality of the estimated speech. On the other hand, when comparing \texttt{Denoise+WPE} to PnPWPE, we observe that the former fails to recover the desired speech from the observed input with the same level of detail. This difference can be observed by examining the solid line rectangles in Fig.~\ref{fig:spec_all}. Based on these findings, we can conclude that our proposed framework, which incorporates deep speech priors, offers a more efficient approach to recovering speech from noisy and reverberant conditions compared to the intuitive concatenation of denoising and dereverberation algorithms.

%\noindent\textbf{Discussion of different DNN-based denoisers:}
%In order to explore the flexibility of PnPWPE, we insert different DNN-based denoisers to our framework. From

\noindent\textbf{Discussion of $V(n,k)$:}
In order to explore the capability of term $V(n,k)$, we conduct a comparison experiment between PnPWPE and PnPWPE without term $V(n,k)$ (termed PnPWPE-without-$V$), where the denoiser used here is FRCRN. Fig.~\ref{fig:spec_nov} and Fig.~\ref{fig:re_nov} present the {bar plot} and the evaluation results respectively. From Fig.~\ref{fig:spec_nov}, it is clear that PnPWPE-without-$V$ fails to remove the additive noise from the observed speech. Moreover, we can see from the {bar plot} in Fig.~\ref{fig:re_nov} that PnPWPE-without-$V$ is able to perform slightly better than the vanilla WPE in most cases, while the performance of PnPWPE improves significantly. For instance, the PESQ value of PnPWPE relatively exceeds that of PnPWPE-without-$V$ by around \textbf{20.6\%}, \textbf{13.3\%}, and \textbf{10.5\%} in the scenarios of WGN, cafe and babble respectively. The same trend can be observed in term of CD and F-SNR. This is strong evidence that introducing the auxiliary term $V(n,k)$ to our framework is reasonable. Further, from this comparison experiments, we can deduce that in addition to incorporating sophisticated speech prior for the framework, the capability of the DNN-based denoiser module in our proposed method is to eliminate some unknown and un-modelled noises, such as structural noise during the optimization processing, instead of additive background noises~\cite{zhang2021plug}.

\noindent\textbf{Convergence property of PnPWPE:}
\begin{figure*}[t!]
\centering
       \includegraphics[width=0.32\textwidth]{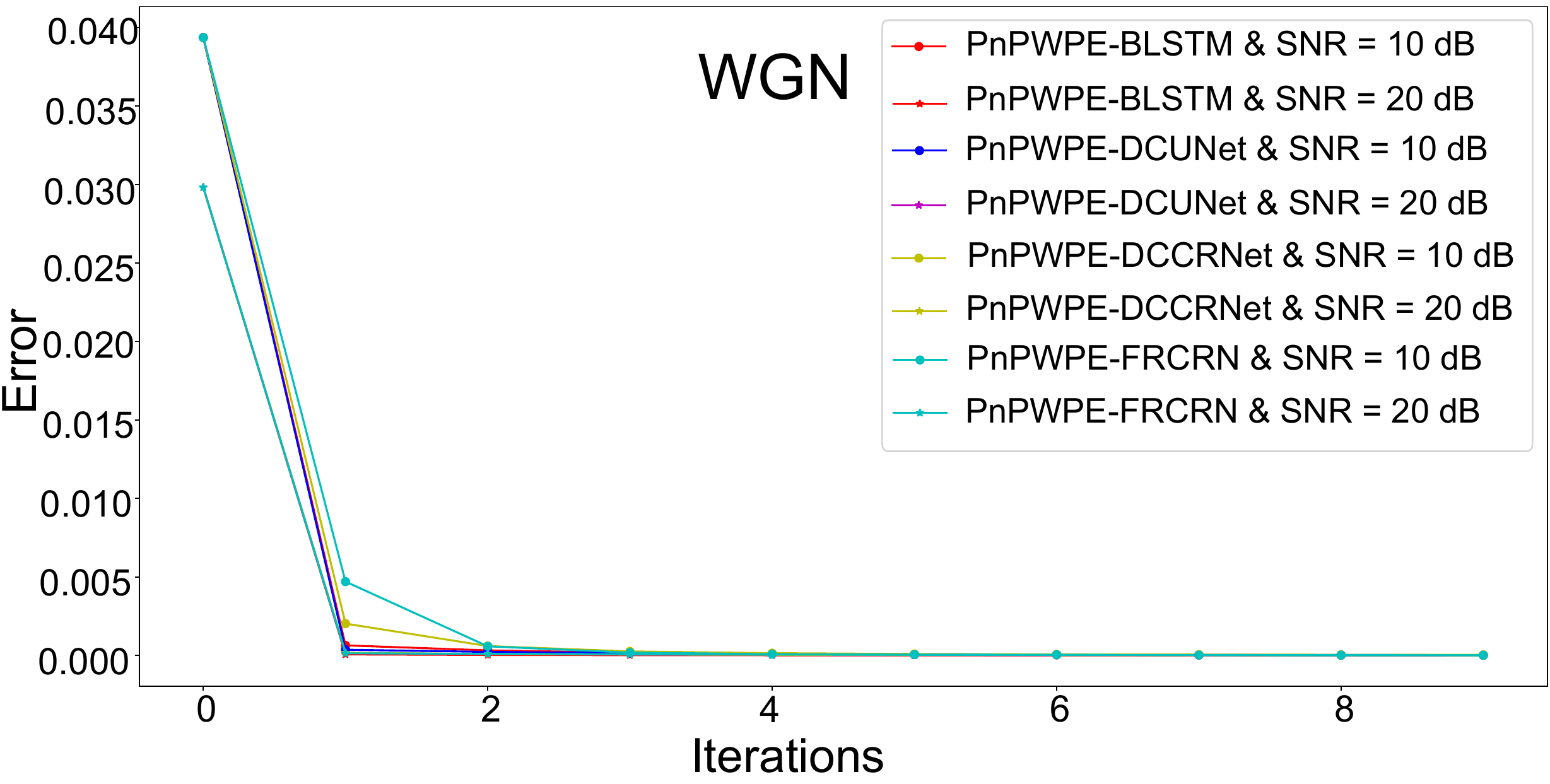}
       \includegraphics[width=0.32\textwidth]{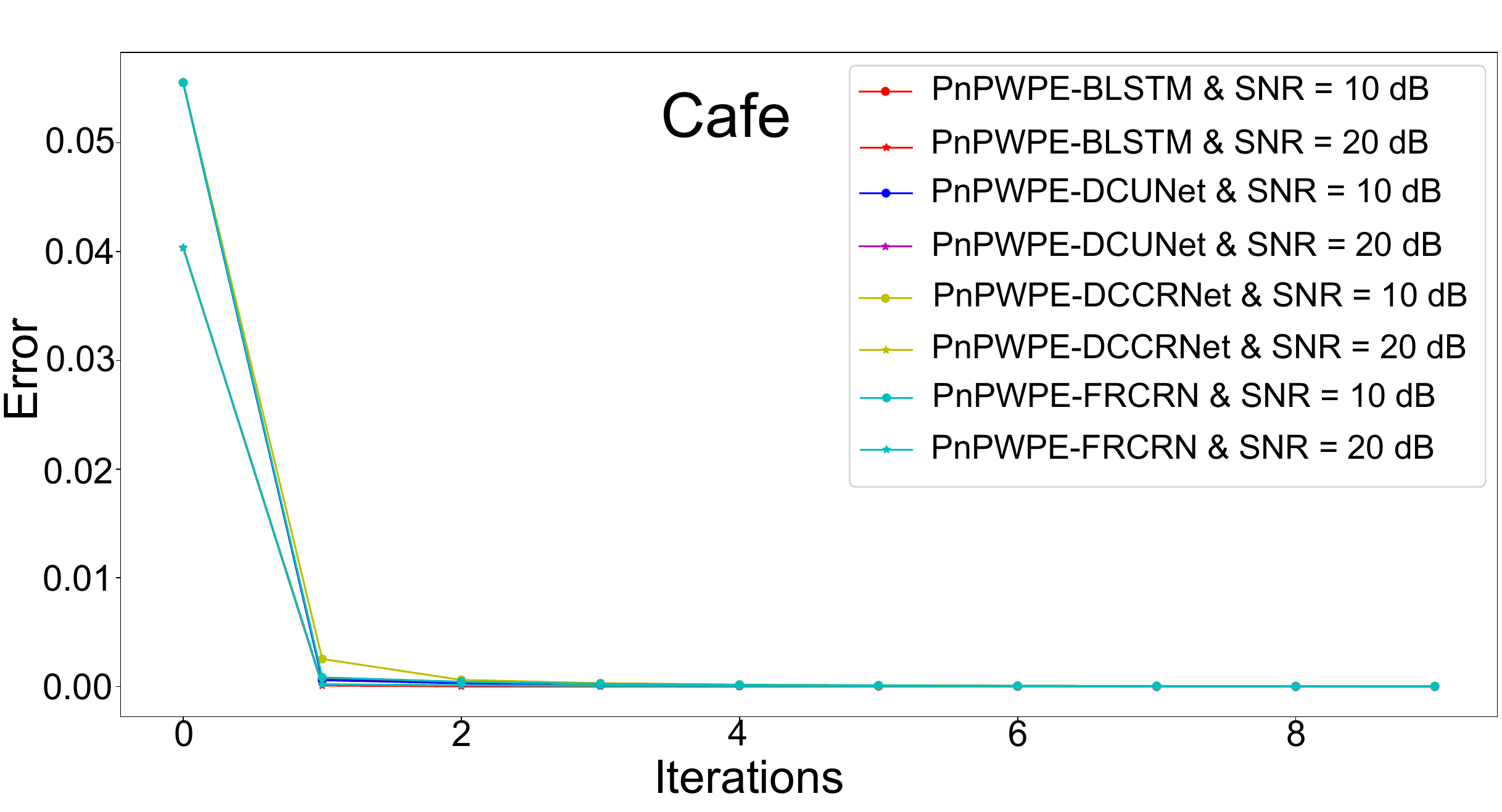}
       \includegraphics[width=0.32\textwidth]{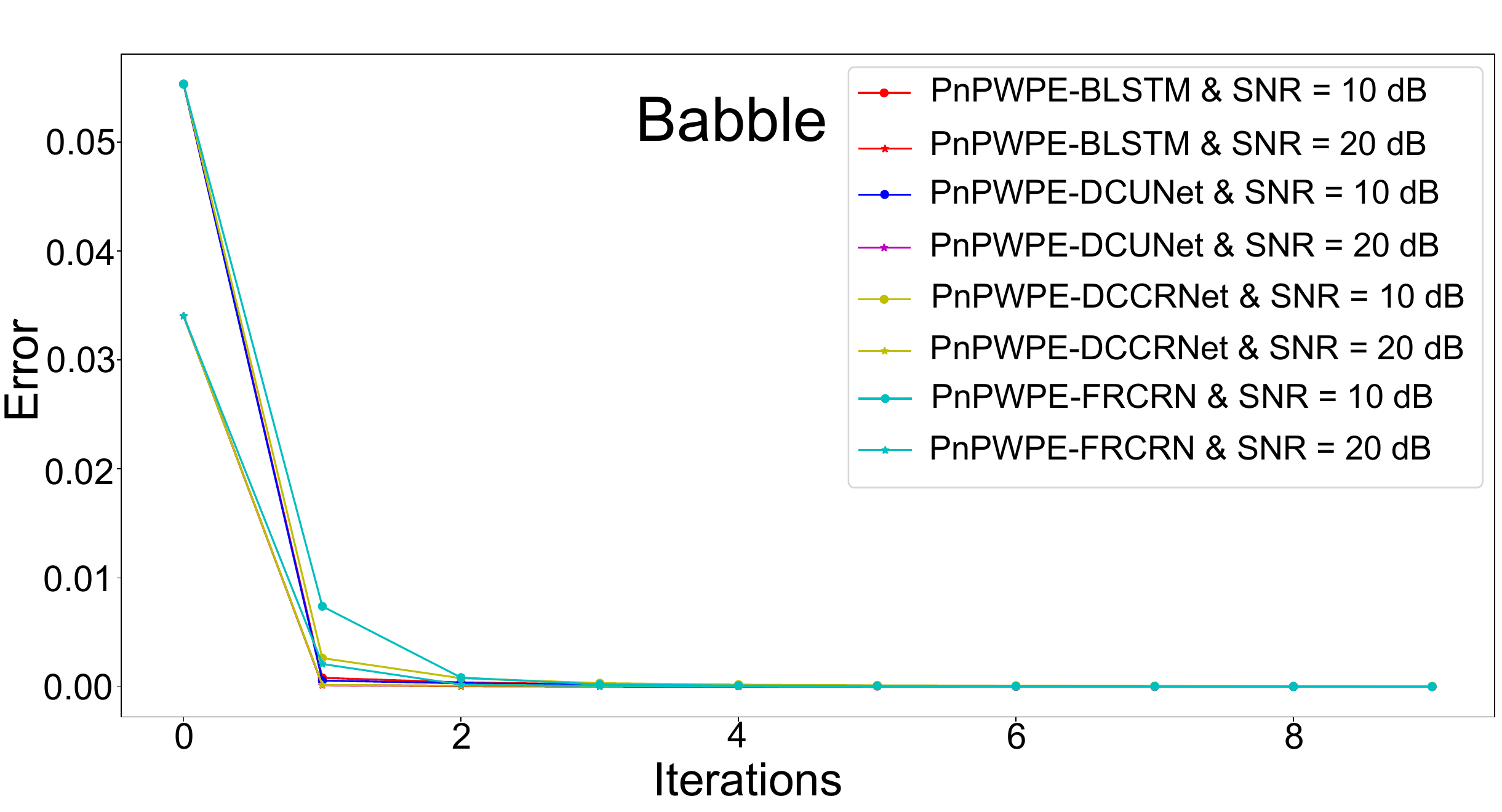}
       \vspace{-2mm}
  \caption{{The Error convergence curves of the proposed framework under the scenarios with all types of noises at SNR=10 dB and 20 dB in Room B.}}
  \label{fig:conv_all}
  \vspace{-5mm}
\end{figure*}
In our work, a DNN-based denoiser is utilized as a black-box to replace the explicitly solution of the subproblem in~\eqref{eq:r}. The theoretical analysis of the convergence can be difficult, as nonlinear denoisers always involve complex operators, such as the BLSTM module. To illustrate the convergence, we present the convergence curves of our proposed framework of all DNN-based denoisers to experimentally show its  convergence property. From Fig.~\ref{fig:conv_all}, we  observe that the proposed framework with each DNN-based denoiser has a stable and robust convergence property in terms of different types of noises. Besides, the proposed framework can reach stability after 3 iterations, demonstrating that an early stop operator can be performed to further save computation time consuming.

\section{Conclusion}
In this paper, we have presented a method for improving the performance and robustness of the WPE algorithm in complex environments with reverberation and additive noise. Our approach integrates data-driven speech priors using a Plug-and-Play (PnP) strategy, specifically employing the RED strategy for speech reconstruction. We implemented various types of DNN-based denoisers and integrated them into the optimization steps, allowing us to effectively capture the prior information from data and demonstrate the flexibility of our proposed framework. The experimental results have confirmed the  performance and robustness of our method under diverse conditions. However, there are several respects for future investigation:
\begin{itemize}
     \item  Online and lightweight implementation: It would be valuable to explore the development of an online and lightweight version of the proposed PnP method. As WPE is commonly employed in online scenarios for practical applications, adapting our framework to real-time settings would enhance its usability.
     \item Alternative integration approaches: Further research can be conducted to investigate alternative ways of integrating data-driven approaches. For instance, instead of using denoisers, inserting a generative module, or employing other strategies for integrating data-driven approaches can be explored like in~\cite{chen2023integration}.
     \item Extension to other speech processing tasks: The combination of physics-based and data-driven approaches holds promise for addressing various speech processing tasks beyond dereverberation. By employing the proposed PnP strategy, other tasks can leverage denoisers to capture speech priors, eliminating the need for training task-dependent networks.
\end{itemize}
\bibliographystyle{IEEEbib}
%\footnotesize

\bibliography{refs}

\vfill

\end{document}